\documentclass[%
 reprint,
superscriptaddress,
amsmath,
amssymb,
aps,
prx,
longbibliography
]{revtex4-2}

\usepackage{hyperref}
\hypersetup{
    colorlinks=true,
    linkcolor=blue,
    filecolor=blue,      
    urlcolor=blue,
    citecolor=blue
    }
\usepackage{braket}
\usepackage{xcolor}
\usepackage{xfrac}
\usepackage{graphicx}
\usepackage{dcolumn}
\usepackage{dsfont}
\usepackage{verbatim}
\usepackage{bm}
\usepackage{makecell}

\begin{document}

\preprint{APS/123-QED}

\title{Quantum simulation with Rydberg ions in a Penning trap}

\author{Wilson S.~Martins}%
\email{W.SantanaMartins@umb.edu}
\affiliation{Institut f{\"u}r Theoretische Physik and Center for Integrated Quantum Science and Technology,
Universit{\"a}t T{\"u}bingen, Auf der Morgenstelle 14, 72076 T{\"u}bingen, Germany}
\affiliation{Department of Physics, University of Massachusetts Boston, Morrissey Boulevard 100, Boston, Massachusetts, 02125, USA}

\author{Markus Hennrich}%
\affiliation{Department of Physics, Stockholm University, SE-106 91 Stockholm, Sweden}

\author{Ferdinand Schmidt-Kaler}%
\affiliation{QUANTUM, Johannes Gutenberg Universit{\"a}t Mainz, Staudingerweg 7, 55128 Mainz, Germany}

\author{Igor Lesanovsky}%
\affiliation{Institut f{\"u}r Theoretische Physik and Center for Integrated Quantum Science and Technology,
Universit{\"a}t T{\"u}bingen, Auf der Morgenstelle 14, 72076 T{\"u}bingen, Germany}
\affiliation{School of Physics and Astronomy, University of Nottingham, Nottingham, NG7 2RD, United Kingdom}
\affiliation{Centre for the Mathematics and Theoretical Physics of Quantum Non-Equilibrium Systems,
University of Nottingham, Nottingham, NG7 2RD, United Kingdom}

\date{\today}

\begin{abstract}
Quantum simulation of interacting many-body spin systems is routinely performed with cold trapped ions, and systems with hundreds of spins have been studied in one and two dimensions.
In the most common realizations of these platforms, spin degrees of freedom are encoded in low-lying electronic levels, and interactions among the spins are mediated through crystal vibrations.
Here we propose a new approach which enables the quantum simulation of two-dimensional spin systems with interaction strengths that are increased by orders of magnitude. This, together with the unprecedented longevity of trapped ions, opens an avenue for the exploration of phenomena that take place on long timescales, e.g., slow and collective relaxation in frustrated and kinetically constrained systems. 
Our platform makes use of the strong dipolar interactions among electronic Rydberg states and planar confinement provided by a Penning trap.
We investigate how the strong electric and magnetic fields that form this trap affect the properties of the Rydberg states and show that spin-spin interaction strengths on the order of MHz are achievable under experimentally realistic conditions.
As a brief illustration of the capabilities of this quantum simulator, we study the entanglement in a frustrated spin system realized by three ions.

\end{abstract}

\maketitle

\newcommand{\e}{\mathrm{e}}


\section{Introduction \label{sec:intro}}

Cold trapped ions arrange into crystalline structures that can mimic condensed-matter systems~\cite{drew_bro_1998, geor_ash_no_2014, noh_ang_2016, alt_bro_2021, fo_pa_po_2024}.
Unlike conventional solids, where interatomic separations are at the subnanometer scale, trapped-ion crystals exhibit interionic distances ranging from a few to tens of micrometers~\cite{thom_2015, gi_john_2025}.
These large spacings enable high-fidelity optical addressing and precise manipulation of the internal electronic states of individual ions~\cite{singer2010, sch_wei_2022, guo_wu_2024, mc_brown_2024}, supporting quantum simulation of interacting spin systems~\cite{deng_po_ci_2005, bla_wi_2008, friedenauer2008, kim2010, Bermudez_2012, bri_saw_2012, boh_saw_2016, gar_boh_2017, zhang2017, monroe2021, Pham_2024}.

In quantum simulation platforms based on trapped ions, spin–spin interactions arise from phonon-mediated optical forces that couple internal electronic states to collective vibrational modes~\cite{porras2004, deng2005, haljan2005, kim2009, wang2013}, leading to typical Ising couplings ranging up to $10\,\text{kHz}$~\cite{hazzard2014, monroe2021}.
Much stronger interactions become accessible when ions are excited to Rydberg states, whose large dipole moments produce strong photon-mediated dipole–dipole couplings~\cite{mu_li_2008, fel_ba_2015, hi_li_fa_2017, hi_po_fa_2017}.
Experiments in Paul traps have already demonstrated MHz-scale interactions and submicrosecond entanglement generation in Rydberg-excited ions~\cite{zhang_po_wei_2020, mok_henn_2020}.

However, the confinement mechanism of Paul traps, which uses inhomogeneous oscillating electric fields, is detrimental to Rydberg excitation --- especially for large two-dimensional (2D) ion crystals~\cite{hi_po_zhan_2019, mar_wil_2025}.
To overcome this limitation, we propose a quantum simulation platform based on Rydberg ions confined in a Penning trap.
Here, a homogeneous magnetic field combined with a static quadrupole electric field ensures confinement in all spatial directions~\cite{bro_ga_1986, Bollinger_1995, hu_bo_1998}.
Penning traps have demonstrated stable trapping of large 2D crystals comprising hundreds of ions under realistic experimental conditions~\cite{ma_go_jo_2013, wan_kei_2013, tan_shan_2021, wol_ph_2024, mc_brown_2024}.
Such crystals provide a robust and scalable environment for quantum simulation, where tunable long-range interactions arise through Rydberg excitation.

In this paper we develop the theory of a quantum simulator based on Rydberg ions in a Penning trap.
Particular attention is given to the fact that involved strong electric and magnetic fields modify the structure of highly excited Rydberg levels.
Under typical trapping conditions, the Rydberg spectrum is located in the so-called Paschen–Back regime, where internal electronic states are characterized by well-defined orbital and spin angular momenta.
In these highly excited states, the diamagnetic interaction becomes significant as the magnetic-field strength increases, inducing mixing between states of different orbital angular momentum.
We analyze how this mixing influences the resulting spin–spin interactions, which are controlled through microwave (MW) dressing of Rydberg levels. 
To illustrate the potential of this platform, we consider a concrete instance of a 2D quantum magnet implemented with a planar three-ion crystal.
In this triangular configuration, the Rabi drive enables access to regimes where the ground state exhibits geometric frustration~\cite{moe_son_chan_2000, moess_son_2001, lie_de_ba_2018}.
For experimentally realistic parameters~\cite{ma_go_jo_2013, goodwin2016}, we find that the resulting dipole–dipole interaction strengths are on the order of MHz. 

These strong interactions, together with the extraordinary robustness and stability of ion crystals, open a new perspective for quantum simulation: the fact that strong confinement is provided for both ground state and Rydberg ions mitigates the impact of mechanical forces \cite{faoro_2016,failache_2025,emperauger_2025}. Furthermore, given that interactions are electrostatic, the vibrational degrees of freedom of trapped Rydberg ions remain available and can be utilized for coherent or dissipative manipulation. This can be exploited for \emph{in situ} laser cooling of the ion crystal in order to prevent heating. Eventually, this may allow to probe dynamical phenomena, such as (quantum) glassy relaxation in many-body systems with kinetic constraints \cite{biroli_2013,le_ga_2013}, which unravel on ultra-long time scales and are inaccessible by current experiments.

\section{Single ion theory}
\label{sec:single_ion}
In this section we develop the theoretical model describing the motional and electronic degrees of freedom of a single Rydberg ion in a Penning trap. 
We discuss how the trap generates confinement of the external motion and analyze the structure of the electronic Rydberg states. 
We also investigate how the electric and magnetic fields of the Penning trap lead to the coupling between the ion's external motion and internal electronic dynamics.

\subsection{Trapping fields and electron-core interaction}
\begin{figure}
    \centering
    \includegraphics[scale=0.44]{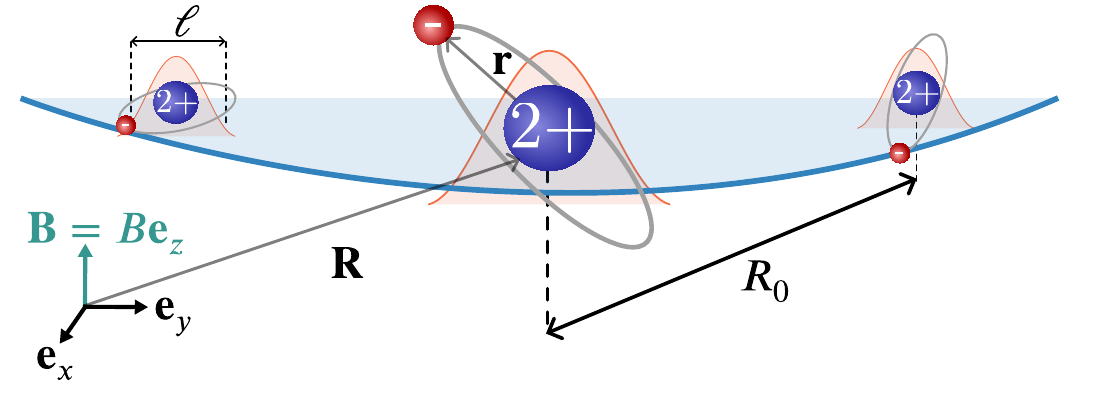}
    \caption{\textbf{Rydberg ions in a Penning trap.}  
    Length scales and coordinate system.
    To describe a single ion, we use $\mathbf{R}$ and $\mathbf{r}$, the center-of-mass and the relative coordinate, respectively. 
    The distance between the ion core and the Rydberg electron scales with the principal quantum number as $\sim n^{2}$, reaching typical values $\langle r \rangle\approx100\, \text{nm}$. 
    For typical trap frequency magnitudes, the harmonic confinement yields an oscillator length of approximately $\ell \approx 10\, \text{nm}$. 
    The equilibrium interparticle spacing, determined by the balance between Coulomb repulsion and harmonic confinement, is on the order of $R_0 \approx 10\,\mu\text{m}$. 
    \label{fig:cartoon}
    }
\end{figure}

We describe a Rydberg ion using a two-body model: the electron in the Rydberg state is treated explicitly, and the ion core, comprising the nucleus and all remaining electrons, is considered as a single particle.
This treatment is analogous to previous works which considered Rydberg ions confined in Paul traps~\cite{mu_li_2008, sch_fel_2011}.
In our analysis, we focus on ionic isotopes with zero nuclear spin, e.g., $^{40}\text{Ca}^{+}$, which is a well-established species in trapped ion experiments~\cite{fel_ba_2015, an_vo_mo_2021}.
Under such assumptions, a single trapped Rydberg ion is well approximated by the minimal coupling Hamiltonian for two charged particles in an electromagnetic field~\cite{lesanovsky2005b}.
The initial Hamiltonian in the laboratory frame reads
\begin{equation}
    \begin{aligned}
        H &= \frac{1}{2 m_{\mathrm{c}}}\big[\mathbf{p}_{\mathrm{c}} - 2e \mathbf{A}(\mathbf{r}_{\mathrm{c}})\big]^{2} + \frac{1}{2 m_{\mathrm{e}}}\big[\mathbf{p}_{\mathrm{e}} + e \mathbf{A}(\mathbf{r}_{\mathrm{e}})\big]^{2} \\
        & -  \boldsymbol{\mu}_{\mathrm{e}}\cdot \mathbf{B} + 2e \phi(\mathbf{r}_{\mathrm{c}}) - e \phi(\mathbf{r}_{\mathrm{e}}) + V(\left|\mathbf{r}_{\mathrm{e}} - \mathbf{r}_{\mathrm{c}}\right|),
        \label{eq:ham_one}
    \end{aligned}
\end{equation}
where $e > 0$ is the elementary charge. 
Here, we have the electronic magnetic moment $\boldsymbol{\mu}_{\mathrm{e}} = -(g_{s}/2m_{\mathrm{e}})\mathbf{s}$, where $g_{s} \approx 2$ and $\mathbf{s}$ is the electronic spin angular momentum; we set $\hbar = 1$ throughout the manuscript.
The electron and the ion core have positions $\mathbf{r}_{\mathrm{e}}$ and $\mathbf{r}_{\mathrm{c}}$, respectively, and $r = \left|\mathbf{r}_{\mathrm{e}} - \mathbf{r}_{\mathrm{c}}\right|$ is their relative distance.
The conjugate momenta for the electron and ion core are $\mathbf{p}_{\mathrm{e}}$ and $\mathbf{p}_{\mathrm{c}}$, respectively.
In addition, the electron has mass $m_{\mathrm{e}}$ and charge $-e$, and the ion core is described by a single particle with mass $m_{\mathrm{c}}$ and charge $+2e$.
A sketch of the model, with relevant spatial quantities and magnetic field orientation, is shown in Fig.~\ref{fig:cartoon}.

The Penning trap features a combination of a homogeneous magnetic field and a quadrupole electric field \cite{bro_ga_1986}.
For the magnetic field, the vector potential is written in the symmetric gauge, defined as $\mathbf{A}(\mathbf{r}) = \frac{1}{2} (\mathbf{B}\times \mathbf{r})$, where the homogeneous magnetic field vector $\mathbf{B} = B \mathbf{e}_{z}$ is oriented along the $z$ direction ($\mathbf{e}_{x}$, $\mathbf{e}_{y}$, and $\mathbf{e}_{z}$ are the unit vectors in cartesian coordinates). 
The potential dictating the interaction of the charged particles with the quadrupole electric field $\boldsymbol{\mathcal{E}}(\mathbf{r})$ can be cast into the form
\begin{equation}
    \phi(\mathbf{r}) = - \boldsymbol{\mathcal{E}}(\mathbf{r})\cdot \mathbf{r} = - \beta(\rho^{2} - 2z^{2}).
\end{equation}
Here, $\rho = \sqrt{x^{2} + y^{2}}$ is the radial coordinate, and $\beta$ denotes the electric field gradient.

The central binding potential $V(r)$ describes the effective interaction between the Rydberg electron and the ionic core. 
It differs from the pure hydrogenic potential as it accounts for the finite size and internal electronic structure of the core charge.
This potential includes contributions from modified Coulomb, polarization, and spin–orbit coupling terms: 
\begin{equation}
    V(r) = V_{\mathrm{c}}(r) + V_{\mathrm{p}}(r) + V_{\mathrm{so}}(r).
\end{equation}
The modified Coulomb potential,
\begin{equation}
    V_{\mathrm{c}}(r) = - \frac{e^{2}}{4\pi \epsilon_{0} r} \big[2 + (Z_{\mathrm{nuc}} - 2) \e^{-\alpha_{l, 1} r} + \alpha_{l, 2} \e^{-\alpha_{l, 3} r}\big],
    \label{eq:corr_potential}
\end{equation}
describes the effective central potential of an ion with core charge $ +2e$.  
Here, $\epsilon_{0}$ is the vacuum permittivity, $Z_{\mathrm{nuc}}$ is the nuclear charge, and $\alpha_{l,i}$ are $l$-dependent parameters tabulated in Ref.~\cite{aymar1996}, where $l$ is the quantum number of orbital angular momentum.
The induced polarization potential,
\begin{equation}
    V_{\mathrm{p}} (r) = - \frac{\alpha_{\mathrm{cp}}e^{2}}{2(4\pi\epsilon_{0})^{2} r^{4}} \Big[1 - \e^{- \left(r/r_{l, \mathrm{c}}\right)^{6}}\Big],
\end{equation}
describes the dipole moment induced in the ionic core by the valence electron. 
Here, $\alpha_{\mathrm{cp}}$ is the static dipole polarizability of the doubly charged ionic core and $r_{l, \mathrm{c}}$ is the cutoff radius, which describes the effective size of the ionic core.
This cutoff is introduced to truncate the unphysical short-range behavior of the polarization potential~\cite{greene1991, marinescu1994}.
Finally, the relativistic spin–orbit coupling,
\begin{equation}
    V_{\mathrm{so}}(r) = \frac{\mathbf{l}\cdot \mathbf{s}}{2m^{2} c^{2} r} \bigg[1 - \frac{V_{l}(r)}{2m_{\mathrm{e}}c^{2}} \bigg]^{-2} \frac{\mathrm{d} V_{l}(r)}{\mathrm{d} r},
    \label{eq:spin_orbit}
\end{equation}
where $V_{l} (r) = V_{\mathrm{c}}(r) + V_{\mathrm{p}} (r)$ amounts for the $l$-dependent, non-relativistic model potential.
The term $[1 - V_{l}(r)/(2m_{\mathrm{e}}c^{2})]^{-2}$ is introduced to regularize the non-physical divergence near the origin~\cite{condon1935}.
Here, $c$ is the speed of light, and $\mathbf{l}$ and $\mathbf{s}$ are the orbital and spin angular momentum operators associated with the internal degrees of freedom, respectively.

\subsection{Rotating center of mass frame and static harmonic confinement}

We introduce the coordinates of the center of mass, $\mathbf{R} = (X, Y, Z)$, and the relative motion, $\mathbf{r} = (x, y, z)$, see Fig.~\ref{fig:cartoon}.
Together with their respective conjugate momenta $\mathbf{P}$ and $\mathbf{p}$, they are defined as
\begin{equation}
\begin{aligned}
    \mathbf{R} &= \frac{m_{\mathrm{c}}\mathbf{r}_{\mathrm{c}} + m_{\mathrm{e}}\mathbf{r}_{\mathrm{e}}}{M}, 
    &\qquad 
    \mathbf{r} &= \mathbf{r}_{\mathrm{e}} - \mathbf{r}_{\mathrm{c}}, \\[4pt]
    \mathbf{P} &= \mathbf{p}_{\mathrm{c}} + \mathbf{p}_{\mathrm{e}}, 
    &\qquad 
    \mathbf{p} &= \frac{m_{\mathrm{c}}\mathbf{p}_{\mathrm{e}} - m_{\mathrm{e}}\mathbf{p}_{\mathrm{c}}}{M}.
\end{aligned}
\label{eq:CM_coord}
\end{equation}
where $M = m_{\mathrm{e}} + m_{\mathrm{c}}$ is the total mass of the system.
Throughout this manuscript, we refer to the center-of-mass motion as the external dynamics and the relative motion as the internal dynamics.
We proceed by defining the unitary transformation~\cite{sch_ce_1991}
\begin{equation}
    U (\mathbf{r}, \mathbf{R}) = \exp\big[\!-\!\mathrm{i} \tfrac{e(m_{\mathrm{c}} + 2m_{\mathrm{e}})}{M}  \mathbf{A}(\mathbf{R}) \cdot \mathbf{r}\big],
\end{equation}
whose application transforms the total Hamiltonian $H$ [in Eq.~\eqref{eq:ham_one}] according to
\begin{equation}
    H \mapsto U^{\dagger} H U =  H_{\mathrm{ex}}(\mathbf{R})+ H_{\mathrm{in}}(\mathbf{r}) + H_{\mathrm{co}}(\mathbf{R}, \mathbf{r}).
    \label{eq:tot_ham}
\end{equation}
Here, we have separated the total Hamiltonian into external, internal, and coupling terms. 
Next, we exploit the fact that ionic core mass is much larger than the electronic mass, $m_{\mathrm{c}} \approx M \gg m_{\mathrm{e}}$, with $M$ being five orders of magnitude larger than $m_{\mathrm{e}}$.
We then arrive at the following expressions for the Hamiltonian terms:
\begin{equation}
    \begin{aligned}
    H_{\mathrm{ex}} &= \frac{1}{2M} \! \Big[\mathbf{P}- \frac{e}{2} \big(\mathbf{B} \times  \mathbf{R}\big)\Big]^2 + e \phi(\mathbf{R}), \\
    H_{\mathrm{in}} &= \frac{1}{2m_{\mathrm{e}}} \!\Big[\mathbf{p}+ \frac{e}{2}\big(\mathbf{B} \times  \mathbf{r}\big)\Big]^2 - \boldsymbol{\mu}_{\mathrm{e}}\cdot \mathbf{B} - e \phi(\mathbf{r}) + V(r), \\
    H_{\mathrm{co}} &= \frac{e}{M} \! \Big[\mathbf{P}- \frac{e}{2}\big(\mathbf{B}\times \mathbf{R}\big)\Big] \!\cdot \!\big(\mathbf{B}\times \mathbf{r}\big) + 2 e \boldsymbol{\mathcal{E}}(\mathbf{r})\cdot \mathbf{R},
\end{aligned}
\label{eq:ham_total}
\end{equation}
where the arguments of the Hamiltonians as shown in Eq.~\eqref{eq:tot_ham} are omitted for readability.
The external Hamiltonian describes a particle with mass $M$ and charge $e$ moving in a homogeneous magnetic and quadrupole electric fields.
The internal Hamiltonian, on the other hand, describes a particle of mass $m_{\mathrm{e}}$ and charge $- e$, also subjected to homogeneous magnetic and quadrupole electric fields, additionally influenced by the central potential $V(r)$.
A detailed derivation of the total Hamiltonian, including the transformations of internal, external, and coupling Hamiltonians, is presented in App.~\ref{app:derivation}.

The energy scales of the external dynamics are dependent on three frequencies, in order named axial, cyclotron, and radial frequencies:
\begin{equation}
    \omega_{z} = \sqrt{\frac{4e\beta}{M}}, \qquad \omega_{\mathrm{c}} = \frac{eB}{M}, \qquad \omega_{\rho} = \frac{1}{2}\sqrt{\omega^{2}_{c} - 2\omega^{2}_{z}}.
    \label{eq:conf_freq}
\end{equation}
Given that the center of mass of the ion rotates with frequency $\omega_{\mathrm{c}}/2$ about the $z$-axis \cite{hu_an_1997}, it is therefore convenient to move into a reference frame that is co-rotating with this cyclotron motion. 
This is achieved via the unitary transformation
\begin{equation}
    U(t) = \exp\big[\!-\!\tfrac{\mathrm{i\omega_{c}}}{2} (L_{z}+ j_{z})t\big],
\end{equation}
Here, $L_{z} = XP_{y} - YP_{x}$ is the $z$ component of the angular momentum of the external motion and $j_{z} = l_{z} + s_{z}$, with $l_{z} = xp_{y} - yp_{x}$, is the $z$ component of the total angular momentum associated with the ion's internal dynamics. 
The introduction of the rotation in spin space, generated by $s_{z}$, guarantees that the central potential is kept invariant under the frame transformation. In applying this transformation according to
\begin{equation}
    H  \mapsto  U H U^{\dagger} + \mathrm{i}  \dot{U}U^{\dagger},
    \label{eq:oxc_frame}
\end{equation}
one finds that the external Hamiltonian becomes
\begin{equation}
    H_{\mathrm{ex}} = \frac{\mathbf{P}^2}{2M} + \frac{M}{2} \big[\omega^{2}_{\rho}(X^{2} + Y^{2}) + \omega^{2}_{z} Z^{2}\big]
    \label{eq:ex_ham}
\end{equation}
which is the Hamiltonian of a 3D harmonic oscillator for the external motion.
Unless stated otherwise, all Hamiltonians from this point onward are expressed in the rotating frame.
According to Eq.~\eqref{eq:conf_freq}, radial confinement is only achieved, i.e., $\omega_{\rho} > 0$, when axial and cyclotron frequencies obey: $\omega_{\mathrm{c}} > \sqrt{2} \omega_{z}$~\cite{bro_ga_1986}.

Experimentally, ion rotation in Penning traps can be stabilized using rotating-wall potentials, which generate periodic electric fields that lock the crystal rotation frequency~\cite{hu_an_1997, hu_bo_1998, mit_bo_1998, powell_2002, to_bri_2016}. 
In the present analysis, we consider the ideal Penning trap potential and do not include additional effects such as rotating-wall fields, magnetic-field inhomogeneities or misalignments, or higher-order electric multipole contributions, thereby keeping the treatment as general as possible. 
These additional terms may lead to small shifts in the normal-mode frequencies; we assume that they can be experimentally controlled to an extent that they do not qualitatively alter the physical picture discussed here.

To conclude this subsection, we provide typical trapping parameters to establish the orders of magnitude of the quantities associated with external confinement. 
For $^{40}\mathrm{Ca}^{+}$ ions, a magnetic field of $B = 1.85~\mathrm{T}$ and an electric field gradient of $\beta = 7.0 \times 10^{5}~\mathrm{V/m^{2}}$ yield axial and cyclotron frequencies of $\omega_{z} = 2\pi \times 412~\mathrm{kHz}$ and $\omega_{\mathrm{c}} = 2\pi \times 707~\mathrm{kHz}$, respectively~\cite{ma_go_jo_2013, goo_stu_2016}. 
For $^{9}\mathrm{Be}^{+}$ ions, with $B = 4.46~\mathrm{T}$ and $\beta = 2.0 \times 10^{6}~\mathrm{V/m^{2}}$, the corresponding frequencies are $\omega_{z} = 2\pi \times 1.47~\mathrm{MHz}$ and $\omega_{\mathrm{c}} = 2\pi \times 7.58~\mathrm{MHz}$~\cite{bri_saw_2012, saw_bri_2014, jor_gil_2019, tan_shan_2021}. 
Other ion species are presented in App.~\ref{app:ion_species}.

\subsection{Electronic states in the presence of electric and magnetic fields} 
\label{sec:mag_effect}

We now investigate how electronic Rydberg states are affected by the electric and magnetic fields of the Penning trap.
The starting point for this analysis is the explicit form of the internal Hamiltonian 
\begin{equation}
    \begin{aligned}
    H_{\mathrm{in}} &= \frac{\mathbf{p}^{2}}{2m_{\mathrm{e}}} + V(r) + \frac{e B}{2m_{\mathrm{e}}} (l_z + g_{s} s_{z}) + \frac{e^{2}B^{2}}{8m_{\mathrm{e}}}\rho^{2}\\
    &+ e \beta (\rho^{2} - 2 z^2).
    \end{aligned}
    \label{eq:int_ham}
\end{equation}
The first and second terms correspond to the kinetic energy and the central potential of the internal electronic motion, respectively.
The third and fourth terms denote Zeeman and diamagnetic couplings, respectively.
The fifth term is the coupling of the electron to the quadrupole electric field.
We base our discussion on the energy eigenstates of the free Hamiltonian
\begin{equation}
    \bigg[\frac{\mathbf{p}^{2}}{2m_{\mathrm{e}}} + V(r)\bigg]\ket{n, l, j, m_j} = E_{nlj}\ket{n, l, j, m_j},
    \label{eq:fine_basis}
\end{equation}
where $E_{nlj}$ is the eigenenergy associated with the eigenstate $\ket{n, l, j, m_j}$, which is degenerate in $m_{j}$.
Here, $j$ and $m_j$ denote the total angular momentum and its projection along the $z$-axis, respectively.
The spin quantum number is omitted since it is fixed at $s = 1/2$.
 
For internal electronic dynamics, the quadrupole electric fields produce energy shifts that are negligible compared to the diamagnetic coupling. 
This follows directly from the scaling of the two terms: while both scale with the internal radial coordinate $\langle \rho^{2} \rangle \sim n^{4}$, the diamagnetic term is proportional to $e^{2}B^{2}/(8m_{\mathrm{e}})$ whereas the quadrupole term is proportional to $e\beta$. 
As a result, the quadrupole interaction becomes comparable to the diamagnetic coupling only for gradients on the order of
\begin{equation}
    \beta_{\max} \approx \frac{e B^{2}}{8m_{\mathrm{e}}} \sim 10^{10}\,\mathrm{V/m^{2}} \times \bigg(\frac{B}{1\,\mathrm{T}}\bigg)^{2}.
\end{equation}
This would exceed by three orders of magnitude the field gradients used in Penning traps; moreover, such strengths are close to the ionization threshold for a Rydberg electron with $n=50$ and $m_{l}=0$, see App.~\ref{app:ionization}.

\begin{figure}
    \centering
    \includegraphics[scale=0.38]{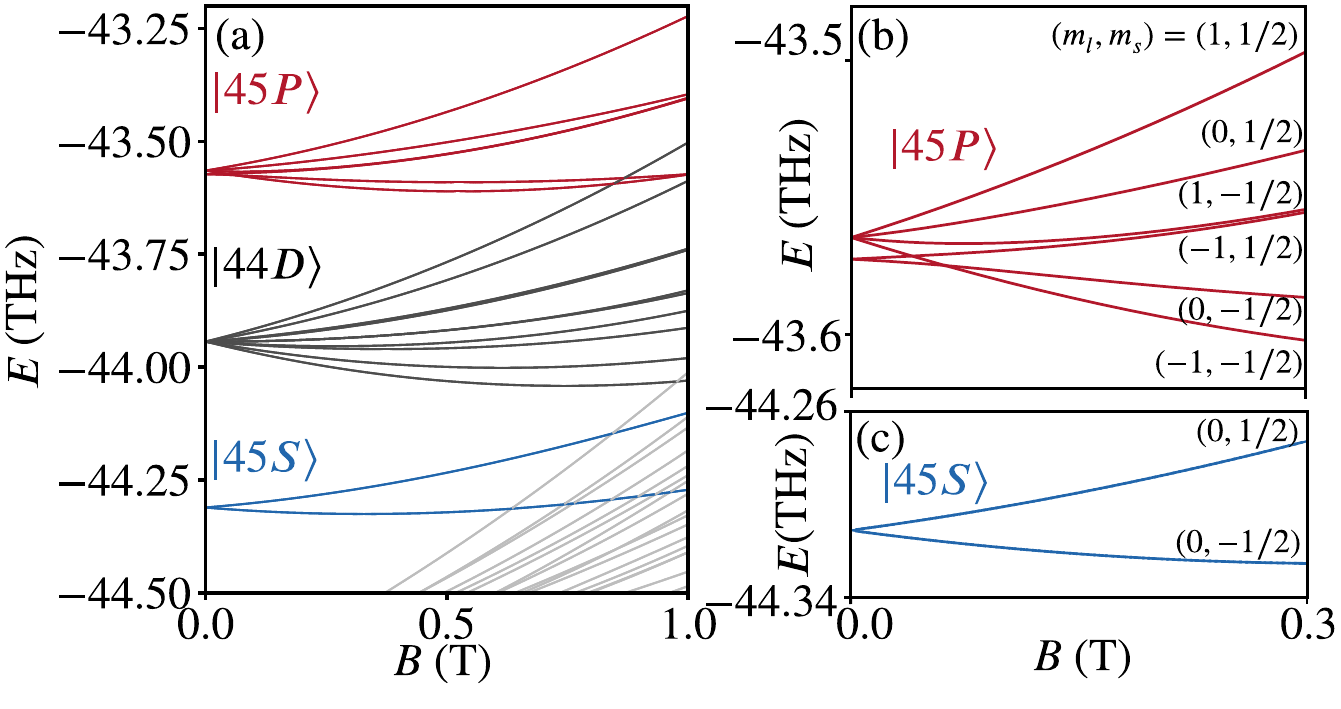}
    \caption{\textbf{Energy spectrum of Rydberg states and the Paschen-Back regime:}
    (a) Spectrum of Rydberg states for $^{40}\text{Ca}^{+}$ ions as a function of the magnetic field $B$. 
    The spectrum exhibits Zeeman splitting and quadratic energy shifts arising from the diamagnetic coupling.
    S, P, and D states are represented in blue, red, and black, respectively; high-angular momentum states ($l > 2$) are plotted in gray.
    (b, c) Magnified view of the Paschen–Back regime for $S$ and $P$ states, where we highlight the dominance of the quantum numbers $m_{s}$ and $m_{l}$.
    These states are used to construct MW-dressed Rydberg states that generate non-vanishing dipole moments.
    The colors are used to highlight the dominant $l$-character of each state, where states with $l>2$ are altogether represented by gray lines.
    The technical approach to obtain the Rydberg energy spectrum here and in the remainder of the manuscript is described in the App.~\ref{app:numerics}.
    }
    \label{fig:pb_regime}
\end{figure}

As the magnetic field $B$ increases, the energies of the Rydberg states split into several components.
In Fig.~\ref{fig:pb_regime}(a), we display states with orbital angular momentum quantum number $l \le 2$, explicitly the $S$, $P$, and $D$ states, and  high-angular momentum states ($l > 2$, i.e., $F$, $G$, and $H$).
For increasing magnetic fields $B$, we distinguish two relevant regimes.
First, in the case of perturbative diamagnetic coupling, the degeneracies associated with $m_{j}$ are lifted, and the Rydberg states evolve into the Paschen–Back regime; see Fig.~\ref{fig:pb_regime}(a), with magnified views of the $S$ and $P$ states shown in Figs.~\ref{fig:pb_regime}(b) and (c). 
In this limit, the internal Hamiltonian obeys the Schrödinger equation, $H_{\mathrm{in}} \ket{\mathbf{L}} = E_{\mathbf{L}} \ket{\mathbf{L}}$, where $\mathbf{L} = \{n, l, m_{l}, m_{s}\}$.
Here, $m_{l}$ and $m_{s}=\pm 1/2$ are magnetic and spin magnetic quantum numbers, respectively.
Second, for stronger diamagnetic coupling, the term $\rho^{2}$ induces quadrupole transitions with $\Delta l = 0, \pm 2$, leading to mixing of Rydberg states with different $l$.

To quantify the $l$-mixing caused by the diamagnetic coupling, we define the magnetic field-dependent states $\ket{E_{\mathbf{L}}} = \ket{E_{\mathbf{L}}(B)}$.
These states are adiabatically connected to the states $\ket{\mathbf{L}}$ in the Paschen-Back regime, and are obtained by solving the Schrödinger equation
\begin{equation}
    H_{\mathrm{in}} (B) \ket{E_{\mathbf{L}}(B)} = E_{\mathbf{L}}(B) \ket{E_{\mathbf{L}}(B)},
    \label{eq:ad_states}
\end{equation}
where the dependency of the internal Hamiltonian on the magnetic field is made explicit. 
For practical purposes, we are mostly interested in states with different orbital angular momentum quantum numbers $l$. 
In such cases, we adopt the shorthand notation $\ket{E_{l}(B)}$ for states adiabatically connected to the field-free state $\ket{l}$ ($l \in \{S, P, D, \ldots\}$, in spectroscopic notation).
Unless explicitly required, the dependence on $B$ will be omitted hereafter for clarity.

Finally, a limiting condition for operating Rydberg ions in a Penning trap is reached when the diamagnetic interaction becomes comparable to the energy spacing between adjacent Rydberg levels.
In this regime, the basis states $\ket{\mathbf{L}}$ reorganize into Landau-like levels.
This threshold can be estimated by comparing the bound-state energy difference of two consecutive principal quantum numbers with the strength of the diamagnetic coupling~\cite{pohl_2009}.
Neglecting spin–orbit coupling and treating the ionic core as point-like, the bound-state energies are given by $E_{n} = -2/(m_{\mathrm{e}}a_{0}^{2}n^{2})$.
The energy difference between neighboring levels is $\Delta E_{n} = E_{n+1} - E_{n}$, which for large $n$ can be approximated as $\Delta E_{n} \approx 4/(m_{\mathrm{e}} a_{0}^{2} n^{3})$.
The expectation value of the squared radial extension is $\langle \rho^{2} \rangle \approx \big(5a_{0}^{2}n^{4}\big)/12$.
Under these assumptions, we obtain the scaling relation between the magnetic field strength $B$ and the principal quantum number $n_{\mathrm{dia}}$ for which the diamagnetic term becomes the dominant contribution to the Hamiltonian:
\begin{equation}
    B > \frac{1}{e a_0^2}\,\sqrt{\frac{384}{5n_{\mathrm{dia}}^{7}}}.
\end{equation}
For $B \ge 2\,\mathrm{T}$, this condition corresponds to $n_{\mathrm{dia}} \gtrsim 52$, signaling the onset of the Landau-level regime~\cite{monteiro1990}.

\subsection{Coupling of internal and external motion}
\label{sec:co_int_ext}

The coupling between the internal and external degrees of freedom leads to state-dependent corrections to both the kinetic energy and the harmonic confinement potential of the external motion. 
These corrections arise from treating the coupling Hamiltonian in the static center-of-mass frame,
\begin{equation} 
    \begin{aligned} 
        H_{\mathrm{co}} &= \frac{eB}{M} \big(xP_{y} - yP_{x}\big) + \bigg(2e\beta - \frac{e^{2}B^{2}}{2M}\bigg)\big(x X + y Y\big) \\ &- 4e\beta z Z,
    \end{aligned} 
\end{equation}
within the Born–Oppenheimer approximation and evaluating it using time-independent perturbation theory~\cite{mu_li_2008, sch_fel_2011}. In this approach, the center-of-mass coordinates $X$, $Y$, and $Z$ are treated as fixed parameters in the internal Hamiltonian, justified by the much faster timescales of internal electronic dynamics compared to those of the vibrational external motion. 
For an internal electronic state $\ket{\mathbf{L}}$ in the Paschen–Back regime, the resulting second-order energy shift takes the form
\begin{equation}
    \begin{aligned}
        \Delta E^{(2)}_{\mathbf L}(\mathbf R,\mathbf P) &=
        \frac{P_x^{2}+P_y^{2}}{2\tilde{M}}
        + \frac{M}{2}\!\left[
        \tilde{\omega}_{\rho}^{2}(X^{2}+Y^{2})
        + \tilde{\omega}_{z}^{2} Z^{2}
        \right] \\
        &\quad
        + \frac{\tilde{\omega}_{\mathrm c}}{2} L_{z}
    + \tilde{\omega}_{\mathrm c}\, YP_{x},
    \end{aligned}
    \label{eq:shift_energy}
\end{equation}
where $\tilde{M}$, $\tilde{\omega}_{\rho}$, $\tilde{\omega}_{z}$, and 
$\tilde{\omega}_{\mathrm c}$ denote the state-dependent modifications to the 
external mass and trapping frequencies induced by the coupling. 
Explicit 
expressions for these quantities, obtained from the perturbative expansion, are given in App.~\ref{app:coup_intext}.

To give a quantitative estimate of the frequency and mass modifications, we focus on the lowest angular momentum Rydberg state $\ket{S} = \ket{n, 0, 0, \pm 1/2}$ and its respective second-order energy correction $\Delta E^{(2)}_{S}$.
The coupling Hamiltonian only causes dipole-type perturbative couplings with states $\ket{P} = \ket{n, 1, m_{l}, \pm 1/2}$, where $m_{l} = 0, \pm 1$.
We further consider only couplings between the states $\ket{S}$ and $\ket{P}$ with the same principal quantum number $n$, which are the energetically closest states.
Under these assumptions, mass and frequency modifications in the Paschen-Back regime can be approximated by the following expressions.
\begin{equation}
    \begin{aligned}
        \tilde{M} &\approx - \frac{3 M^{2}}{e^{2}B^{2}} \frac{E_{P} - E_{S}}{\left|\bra{S} r \ket{P}\right|^{2}},\\
        \tilde{\omega}^{2}_{\rho}&\approx - \frac{1}{3 M}\bigg(2e\beta - \frac{e^{2}B^{2}}{2M}\bigg)^{2}\frac{\left|\bra{S} r \ket{P}\right|^{2}}{E_{P} - E_{S}},\\
        \tilde{\omega}^{2}_{z} &\approx - \frac{32e^{2}\beta^{2}}{3M}\frac{\left|\bra{S} r \ket{P}\right|^{2}}{E_{P} - E_{S}},\\
        \tilde{\omega}_\mathrm{c} &\approx -\frac{2eB}{3M} \bigg(4 e \beta - \frac{e^{2}B^{2}}{M}\bigg) \frac{\left|\bra{S} r \ket{P}\right|^{2}}{E_{P} - E_{S}}.
    \end{aligned}
    \label{eq:corr_shifts}
\end{equation}
Here, we have ignored the energy splitting for states $\ket{P}$ with different $m_{l}$.
The modified axial and radial frequencies resulting from the internal-external coupling for Rydberg-excited ions are 
\begin{equation}
    \omega'_{\rho, z} = \omega_{\rho, z}\sqrt{1 + \frac{\tilde{\omega}^{2}_{\rho, z}}{\omega^{2}_{\rho, z}}} \approx \omega_{\rho, z} + \frac{\tilde{\omega}^{2}_{\rho, z}}{2\omega_{\rho, z}}.
\end{equation}
The frequency shifts for axial and radial confinement are $\Delta \omega_{\rho, z} = \tilde{\omega}^{2}_{\rho, z}/(2\omega_{\rho, z})$. 
Furthermore, the shifts associated with the cyclotron frequency and external mass are $\Delta \omega_{\mathrm{c}} = \tilde{\omega}_{\mathrm{c}}$ and $\Delta M = M^{2}/(M + \tilde{M})$.
For magnetic fields $B \le 2\,\mathrm{T}$, the relative corrections $\Delta \omega_{\rho, z}/\omega_{\rho, z}$, $\Delta\omega_{\mathrm{c}}/ \omega_{\mathrm{c}}$, and $\Delta M/M$ remain below $10^{-3}$ for principal quantum numbers from $30$ to $50$.

\section{Interacting Rydberg ions} 
\label{sec:int_ryd}

In this section, we investigate systems composed of several trapped Rydberg ions.
We begin by expanding the electrostatic potential between ions into multipolar contributions, which reveal the hierarchy of electrostatic couplings (charge-charge, dipole–charge, dipole–dipole, and quadrupole-charge).
These terms are then collected into internal, external, and coupling many-body Hamiltonians.
Finally, we show how MW-dressing allows for the generation of strong dipole-dipole interactions between ionic Rydberg states.

\subsection{Electrostatic interaction between trapped Rydberg ions}

The Hamiltonian for $N$ interacting trapped Rydberg ions has the form
\begin{equation}
    \mathcal{H} = \sum^{N}_{i = 1}  H_{i} + \frac{1}{2}\sum^{N}_{\substack{i,j=1\\i \ne j}}V_{ij} = \mathcal{H}_{\mathrm{in}} + \mathcal{H}_{\mathrm{ex}} + \mathcal{H}_{\mathrm{co}},
    \label{eq:many_body}
\end{equation}
where, similarly to Eq.~\eqref{eq:tot_ham}, we have separated the Hamiltonian in terms of external, internal, and coupled dynamics of interacting Rydberg ions.
In this equation, the first sum collects single-particle Hamiltonians described in Eq.~\eqref{eq:ham_total} and transformed into the rotating frame in Eq.~\eqref{eq:oxc_frame}.
The second sum describes the electrostatic interactions between ions $i$ and $j$, 
\begin{equation}
    \begin{aligned}
        V_{ij} &= \frac{e^{2}}{4\pi\epsilon_{0}}\!\bigg[\frac{4}{\left|\mathbf{R}_{i} - \mathbf{R}_{j}\right|} \!-\! \frac{2}{\left|\mathbf{R}_{i} \!-\!(\mathbf{R}_{j} + \mathbf{r}_{j})\right|}  \\
        &\qquad \qquad \!-\! \frac{2}{\left|(\mathbf{R}_{i} + \mathbf{r}_{i})\!- \! \mathbf{R}_{j}\right|} \!+\! \frac{1}{\left|(\mathbf{R}_{i} + \mathbf{r}_{i}) \!- \!(\mathbf{R}_{j} + \mathbf{r}_{j})\right|}\bigg].
    \end{aligned}
    \label{eq:elec_pot}
\end{equation}
We consider a regime in which the spatial extension of the Rydberg-excited ion wavefunctions is much smaller than the interparticle distances: 
\begin{equation}
    \langle \mathbf{r}_{i} \rangle, \langle \mathbf{r}_{j} \rangle \sim \langle r \rangle \ll \langle \mathbf{R}_{i} - \mathbf{R}_{j} \rangle \sim R_{0}.
\end{equation}
A sketch with these lengths and their respective scales is shown in Fig.~\ref{fig:cartoon}.
We then expand the potential $V_{ij}$ in electric multipoles, and truncate the expansion at second order in $\langle r \rangle/R_{0}$.
This results in the expression
\begin{equation}
        \begin{aligned}
        V_{ij} &= \frac{e^{2}}{4\pi \epsilon_{0}}\bigg[\frac{1}{R_{ij}} + \frac{\mathbf{n}_{ij}\cdot (\mathbf{r}_{j} - \mathbf{r}_{i})}{ R_{ij}^2} \\
        & \qquad \qquad+ \frac{\mathbf{r}^{2}_{i} - 3(\mathbf{n}_{ij}\cdot \mathbf{r}_{i})^2 + \mathbf{r}^{2}_{j} - 3(\mathbf{n}_{ij}\cdot \mathbf{r}_{j})^2}{2 R_{ij}^3} \\
        & \qquad \qquad + \frac{\mathbf{r}_{i} \cdot \mathbf{r}_{j} - 3(\mathbf{n}_{ij}\cdot \mathbf{r}_{i}) (\mathbf{n}_{ij}\cdot \mathbf{r}_{j})}{R_{ij}^3}\bigg], 
        \end{aligned}
        \label{eq:elec_expand}
\end{equation}
where we have defined the distances between the center of mass positions $R_{ij} = \left|\mathbf{R}_{i} - \mathbf{R}_{j}\right|$, and the normalized axis vector between the pair of ions $i$ and $j$, $\mathbf{n}_{ij} = (\mathbf{R}_{i} - \mathbf{R}_{j})/R_{ij}$.
The first term in the expansion represents the Coulomb interaction between two singly charged ions and is independent of their electronic excitation.
For Rydberg ions, the displacement of the valence electron from the ionic core induces an electric dipole moment --- which interacts with the charge of the other ion ---, giving rise to the second term, the charge-dipole interaction.
The third term describes the interaction between the charge and the quadrupole moment of the Rydberg electron distribution.
Both second and third terms are zero for neutral Rydberg atoms, which can be shown by retaining the core charge in the multipole expansion~\cite{wilkinson2025}.
Finally, the fourth term corresponds to the dipole–dipole interaction, which becomes particularly significant when both ions are excited to Rydberg states.
Each of these interactions contributes a distinct electrostatic effect, which is analyzed in the following.

At sufficiently low temperatures, the ions crystallize into well-defined structures called Coulomb crystals~\cite{bollinger1994}.
In such ion crystals, the ions vibrate about their equilibrium position, and their stability results from the balance of repulsive electrostatic forces between the ions and the trap confinement.
Explicitly, the equilibrium positions of the center of mass of the crystal ions, $\mathbf{R}^{0}_{i}$, are calculated by solving the set of equations for each ion~\cite{james1998}
\begin{equation}
    \nabla_{i} V_{\mathrm{ex}}|_{\mathbf{R}_{i} = \mathbf{R}^{0}_{i}} = 0.
\end{equation}
The external potential $V_{\mathrm{ex}}$ that dictates the center-of-mass motion is defined as
\begin{equation}
    V_{\mathrm{ex}} = \sum^{N}_{i = 1} \frac{M}{2}\big[\omega^{2}_{\rho} \big(X^{2}_{i} + Y^{2}_{i}\big) + \omega^{2}_{z}Z^{2}_{i}\big] + \frac{1}{2}\frac{e^{2}}{4\pi \epsilon_{0}}\sum^{N}_{\substack{i,j=1\\i \ne j}} \frac{1}{R_{ij}}.
    \label{eq:ext_pot}
\end{equation}
The ratio between the axial and radial confinement governs the geometry of the ion crystal.
We define the aspect ratio 
$\alpha = \omega^{2}_{z} / \omega^{2}_{\rho}$~\cite{schiffer1993}. 
For small values of $\alpha$, the ions arrange in a linear string along the weakly confined direction, whereas increasing $\alpha$ favors configurations that are flattened into the transverse $xy$-plane. 
In practice, planar crystals already appear for $\alpha$ of order unity and do not require the asymptotic limit $\omega_{z} \gg \omega_{\rho}$; the precise crossover value $\alpha_{c} = \alpha_{c} (N)$ depends on the number of ions in the crystal~\cite{dubin1993}. 
Thus, throughout this work, planar confinement refers to the regime  $\alpha \gtrsim \alpha_{c}$, in which out-of-plane distortions are suppressed and the crystal remains effectively two-dimensional~\cite{du_on_1999}.

The Hamiltonian governing the external vibrational motion of $N$ ions becomes
\begin{equation}
    \mathcal{H}_{\mathrm{ex}} = \sum^{N}_{i = 1} \frac{\mathbf{P}^{2}_{i}}{2M} + V_{\mathrm{ex}}.
\end{equation}
By performing a harmonic expansion in small displacements around the equilibrium positions $\mathbf{R}^{0}_{i}$, the external Hamiltonian can be written in terms of vibrational normal modes~\cite{ja_1998}, resulting in the collection of harmonic oscillators:
\begin{equation}
    \mathcal{H}_{\mathrm{ex}} = \sum^{3N}_{\alpha = 1} \omega_{\alpha} \big(a^{\dagger}_{\alpha}a_{\alpha} + 1/2\big).
\end{equation}
Here, $a^{\dagger}_{\alpha}$ and $a_{\alpha}$ are creation and annihilation operators associated with the vibrational normal mode of each degree of freedom, where $\alpha = 1, ..., 3N$.
In App.~\ref{app:normal_modes}, we explicitly derive the normal modes of vibration for a planar three-ion crystal. 

\begin{figure}
    \centering
    \includegraphics[scale=0.65]{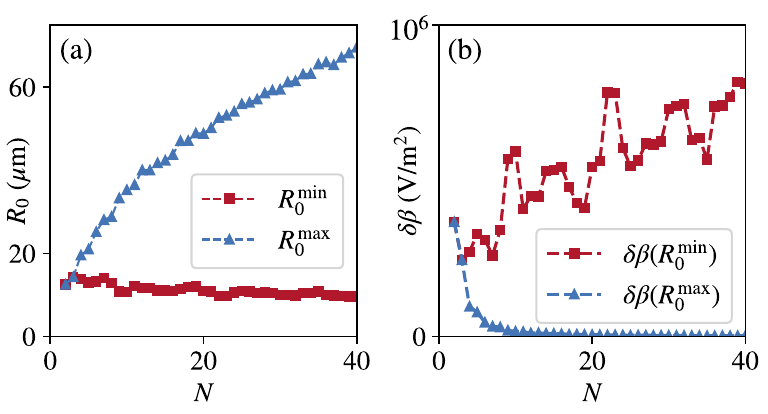}
    \caption{\textbf{Gradient shifts induced by charge--quadrupole coupling in a planar ion crystal.}
    (a) Minimum and maximum inter-ion distances, denoted by $R^{\mathrm{min}}_{0}$ and  $R^{\mathrm{max}}_{0}$, respectively, as functions of the ion number $N$. 
    Here, $R^{\mathrm{min}}_{0}$ corresponds to the shortest separation between any pair of ions in the crystal, while $R^{\mathrm{max}}_{0}$ denotes the largest inter-ion distance.
    (b) Corresponding electric-field-gradient shifts $\delta \beta$ generated by the charge--quadrupole interaction as a function of $N$. 
    The largest shift occurs at the minimum inter-ion distances and reaches a value of approximately $\delta \beta \le 8 \times 10^{5}\, \text{V/m}^{2}$.
    We set $\omega_{\rho} = 2\pi \times 300\,\text{kHz}$, assuming sufficiently strong axial confinement such that the ions form a planar crystal. 
    For $N = 40$, an axial trapping frequency $\omega_{z} = 2\pi \times 900\, \text{kHz}$ ensures planar confinement for all ions. 
    The corresponding trap parameters are $\beta = 3.3 \times 10^{6}\, \text{V/m}^{2}$ and a magnetic field $B = 2.83 \, \mathrm{T}$.}
    \label{fig:shift_grad}
\end{figure}

We proceed by describing internal and coupling Hamiltonians for $N$ trapped Rydberg ions.
We define charge-dipole, dipole-dipole, and charge-quadrupole Hamiltonians:
\begin{subequations}
    \begin{align}
    \mathcal{H}_{\mathrm{cd}} &= \frac{1}{2} \frac{e^{2}}{4\pi\epsilon_{0}} \sum^{N}_{\substack{i,j=1\\i \ne j}} \bigg[\frac{\mathbf{n}_{ij}\cdot (\mathbf{r}_{j} - \mathbf{r}_{i})}{ R_{ij}^2} \bigg] , \\
    \mathcal{H}_{\mathrm{dd}} &= \frac{1}{2} \frac{e^{2}}{4\pi \epsilon_{0}} \sum^{N}_{\substack{i,j=1\\i \ne j}}\bigg[\frac{\mathbf{r}_{i} \cdot \mathbf{r}_{j} - 3(\mathbf{n}_{ij}\cdot \mathbf{r}_{i}) (\mathbf{n}_{ij}\cdot \mathbf{r}_{j})}{R_{ij}^3}\bigg],\\
    \mathcal{H}_{\mathrm{cq}} &= \frac{1}{2} \frac{e^{2}}{4\pi \epsilon_{0}} \sum^{N}_{\substack{i,j=1\\i \ne j}}\bigg[\frac{\mathbf{r}^{2}_{i} - 3(\mathbf{n}_{ij}\cdot \mathbf{r}_{i})^{2}}{2R^{3}_{ij}}\bigg].
\end{align}
\end{subequations}
We first consider the internal electronic motion, which describes the relative motion of the valence electrons in the modified central potential, superimposed by the electric and magnetic potentials of the Penning trap, and other valence electrons. 
The internal Hamiltonian then results in
\begin{equation}
    \begin{aligned}
    \mathcal{H}_{\mathrm{in}} &= \sum^{N}_{i = 1} \bigg[\frac{\mathbf{p}^{2}_{i}}{2m_{\mathrm{e}}} + V(r_i) \bigg] + \frac{e B}{2m_{\mathrm{e}}} \sum^{N}_{i = 1} (l_{z, i} + g_{s} s_{z, i})\\
    & +\!\frac{e^{2}B^{2}}{8m_{\mathrm{e}}}\sum^{N}_{i = 1} \rho^{2}_{i} + e \beta \sum^{N}_{i = 1}  (\rho^{2}_{i}-2z^{2}_{i}) \!+\! \mathcal{H}_{\mathrm{cq}} \!+\! \mathcal{H}_{\mathrm{dd}}.
    \end{aligned}
\end{equation}
The charge-quadrupole term $\mathcal{H}_{\mathrm{cq}}$ generates effective gradient shifts whose intensity can be estimated as $\delta \beta = e/(8 \pi \epsilon_{0} R^{3}_0)$. 
For inter-ion center of mass distances of $R_{0} = 10\,\mu\text{m}$, quadrupole-charge coupling generates electric field gradients of $\delta \beta \approx 10^{6}\,\text{V/m}^{2}$. 
As for the quadrupole electric potential, the effective shifts due to quadrupole-charge coupling are negligible compared to the scales of the magnetic field for intensities considered here; electric-field contributions only become comparable to magnetic effects for $\beta \approx 10^{10}\,\text{V/m}^{2}$ when $B = 1\,\text{T}$.
We estimate maximum and minimum electric-field-gradient shifts due to charge--quadrupole coupling as a function of the number of ions in the trap; for typical trap parameters, see Fig.~\ref{fig:shift_grad}.
Our findings confirm that gradient shifts are negligible in the considered parameter regimes for different crystal sizes.

In the Paschen-Back regime, we then include the quadrupole-charge contribution into the quadrupole electric field potential and approximate the internal Hamiltonian as
\begin{equation}
    \mathcal{H}_{\mathrm{in}} \approx \sum^{N}_{i = 1} \sum_{\mathbf{L}} E_{\mathbf{L}} \ket{\mathbf{L}}\!\bra{\mathbf{L}}_{i} + H_{\mathrm{dd}}.
\end{equation}

Finally, we turn to the coupling between external vibrational motion and internal electronic states.
This term corresponds to the coupled dynamics of a single trapped ion, arising from the coupling of the ions' charge motion with the Penning trap's electric and magnetic fields and the dipolar terms of the multipole expansion of the interaction potential.
The coupling Hamiltonian thus reads
\begin{equation}
    \begin{aligned}
        \mathcal{H}_{\mathrm{co}} &= \frac{eB}{M} \!\sum^{N}_{i = 1}\big(x_{i}P_{y, i} - y_{i}P_{x, i}\big) \!-\! \frac{e^{2}B^{2}}{2M}\sum^{N}_{i = 1}\big(x_{i} X_{i}\!+\!y_{i} Y_{i}\big) \\
        & + 2e\beta \sum^{N}_{i = 1}\big(\mathbf{R}_{i}\cdot\mathbf{r}_{i} - 3Z_{i}z_{i}\big) + H_{\mathrm{cd}}.
    \end{aligned}
\end{equation}
We have established all the ingredients to describe the collective physics of trapped Rydberg ions.

\subsection{Interaction between microwave dressed Rydberg ions}
\label{subsec:mw_states}

In the following, we implement a scheme for generating strong interactions between ions with MW-dressed Rydberg states.
For bare Rydberg states, the dipole moments vanish, and the resulting van der Waals interactions are small for Rydberg ions compared to neutral Rydberg atom interactions.
We thus generate strong interactions via (near)-resonant MW dressing to couple the internal electronic $S$ and $P$ states~\cite{zhang_po_wei_2020, Bao_2025}.
This choice is motivated by the fact that these states are readily accessible via laser excitation from the electronic ground state and are energetically well separated from the degenerate manifold of higher-angular-momentum states, see Fig.~\ref{fig:pb_regime}.

In the setting considered here, we assume that both individual addressing with laser and MW fields and state-selective readout can be achieved. 
Since the ion crystal undergoes a global rotation, the fluorescence image of each ion traces a circular trajectory on the camera. 
To avoid blurring between neighbouring ions, one may employ sufficiently short exposure times, use stroboscopic illumination synchronized to the crystal rotation period, or utilize time-resolved imaging systems. 
Such techniques have already been demonstrated experimentally~\cite{mc_brown_2024, makadia_2026}.
For the trap parameters considered here, typical inter-ion separations are on the order of $10\,\mu\mathrm{m}$, which considerably relaxes the requirements on optical focusing for both addressing and detection. 
In addition, the use of a permanent-magnet Halbach assembly can provide improved optical access, allowing laser beams to be introduced along the $x$, $y$, and $z$ directions.

Below, we describe a bichromatic oscillating field that generates MW-dressing of Rydberg states and laser excitation.
In the dipole approximation, this field takes the form $\boldsymbol{\mathcal{E}}_{\mathrm{field}}(t) = \mathcal{E}_{\mathrm{field}}(t)\mathbf{e}_z$, where
\begin{equation}
    \mathcal{E}_{\mathrm{field}}(t) = \mathcal{E}_{\mathrm{L}}\cos(\omega_{\mathrm{L}}t) + \mathcal{E}_{\mathrm{MW}}\cos(\omega_{\mathrm{MW}}t),
\end{equation}
and $\omega_{\mathrm{L}}$ ($\omega_{\mathrm{MW}}$) and $\mathcal{E}_{\mathrm{L}}$ ($\mathcal{E}_{\mathrm{MW}}$) denote the laser (MW) frequencies and field amplitudes, respectively.
The corresponding Hamiltonian describing the coupling of the oscillating field with the internal motion reads
\begin{equation}
    \mathcal{H}_{\mathrm{field}}(t) = -e \mathcal{E}_{\mathrm{field}}(t) \sum^{N}_{i = 1} z_{i}.
\end{equation}
First, we note that this oscillating field Hamiltonian is invariant under the frame transformation in Eq.~\eqref{eq:oxc_frame}.
Furthermore, this oscillating field drives transitions between states with orbital angular momentum quantum number $\Delta l = \pm 1$, conserving the magnetic quantum number $m_l$.
To include the oscillating field, we redefine the internal Hamiltonian as
\begin{equation}
    \mathcal{H}_{\mathrm{in}} \rightarrow \mathcal{H}_{\mathrm{in}}' = \mathcal{H}_{\mathrm{in}} + \mathcal{H}_{\mathrm{field}}.
\end{equation}
The oscillating field also generates an external motion term proportional to $eZ\mathcal{E}_{\mathrm{field}}(t)$.
This term can be neglected since the laser and MW frequencies, on the order of at least a few GHz, do not couple to the external dynamics.

The MW field couples the Rydberg states
\begin{equation}
    \begin{aligned}
    \ket{S} &\equiv \ket{n, 0, 0, \pm 1/2}, \\
    \ket{P} &\equiv \ket{n, 1, 0, \pm1/2}, 
    \label{eq:states}
\end{aligned}
\end{equation}
In addition, the laser field couples the MW-dressed Rydberg states with the ground state $\ket{G}$, which depends on the specifics of the excitation scheme.
As an example, in experiments with $^{40}\mathrm{Ca}^{+}$ ions, the Rydberg excitation was implemented through a two-photon process~\cite{an_vo_mo_2021, bao_vo_2025}. 
In this case, the ground state of the valence electron is $\ket{4S_{1/2}} = \ket{4,0,0,\pm 1/2}$, and couples to the Rydberg manifold through the intermediate states $\ket{3D_{5/2}}$ and $\ket{5P_{3/2}}$, leading to either $\ket{nS_{1/2}}$ or $\ket{nD_{5/2}}$ depending on whether the targeted Rydberg state is of $S$ or $D$ type.
Under $\pi$-polarized excitation and choosing the Rydberg state $\ket{S} = \ket{n, 0, 0, \pm 1/2}$ for the excitation target, the selection rule $\Delta m_l = 0$ enforces the two-photon pathway $\ket{3,2,0,\pm 1/2} \to \ket{5,1,0,\pm 1/2} \to \ket{S}$.
Rydberg states can be accessed via the transition $\ket{5,1,0,\pm 1/2} \to \ket{S}$ with a wavelength of $286\,\text{nm}$~\cite{bao_vo_2025}.
In App.~\ref{app:ion_species}, Tab.~\ref{tab:ion} and Tab.~\ref{tab:2ph} show masses and one- and two-photon excitation schemes for different ion species which can be used for Rydberg excitation.

\begin{figure}
    \centering
    \includegraphics[scale=0.38]{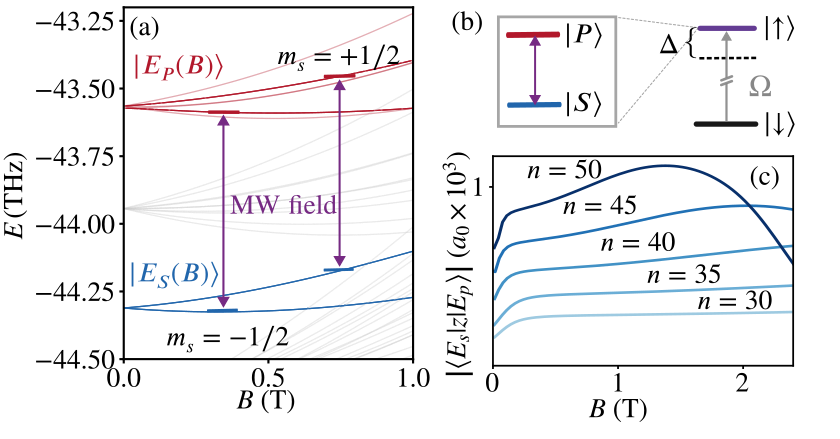}
    \caption{\textbf{Microwave dressing scheme and relevant dipole matrix element:}
    (a) Spectrum of Rydberg states for $^{40}\text{Ca}^{+}$ ions as a function of the magnetic field strength $B$ for principal quantum number $n = 45$. 
    Here, we highlight the states $S$ and $P$, which are combined to form the MW-dressed Rydberg states.
    The purple lines indicate the coupling between selected states, where each of them corresponds to a spin magnetic quantum number $m_{s} = \pm 1/2$.
    (b) Two-level system consisting of the ground state $\ket{\downarrow}$ and the dressed Rydberg state $\ket{\uparrow}$. 
    (c) Dipole matrix element of Rydberg states as a function of the magnetic field strength $B$, for $m_{s} = -1/2$.
    Here, we highlight states $S$ and $P$, represented in blue and red, respectively.}
    \label{fig:mw_scheme}
\end{figure}

We apply the unitary transformation to move into a rotating frame with respect to both MW and laser fields 
\begin{equation}
    \mathcal{H}_{\mathrm{in}} \rightarrow U \mathcal{H}_{\mathrm{in}} U^{\dagger} + \mathrm{i} \dot{U}U^{\dagger},   
\end{equation}
where
\begin{equation}
    \begin{aligned}
        U(t) &= \bigotimes^{N}_{i=1}\e^{\mathrm{i}E_{G}t} \big[e^{\mathrm{i} (\omega_{\mathrm{L}} + \omega_{\mathrm{MW}})t}\ket{P}\!\bra{P}_{i} \\
        &\qquad \qquad \qquad \qquad + e^{\mathrm{i} \omega_{\mathrm{L}}t}\ket{S}\!\bra{S}_{i} + \ket{G}\!\bra{G}_{i}\big].
    \end{aligned}
\end{equation}
Upon applying the rotating-wave approximation to eliminate the fast oscillating terms, the internal Hamiltonian becomes
\begin{equation}
    \begin{aligned}
        \mathcal{H}_{\mathrm{in}} &= \sum^{N}_{i = 1}\big[\Delta_{\mathrm{L}} \ket{S}\!\bra{S}_{i} + (\Delta_{\mathrm{L}} + \Delta_{\mathrm{MW}})\ket{P}\!\bra{P}_{i}\big] + H_{\mathrm{dd}} \\
        &-\!\sum^{N}_{i = 1}\Big[ \frac{\Omega_{\mathrm{L}}}{2} (\ket{G}\!\bra{S}_{i} + \mathrm{H.c.}) \!+\! \frac{\Omega_{\mathrm{MW}}}{2}(\ket{S}\!\bra{P}_{i} + \mathrm{H.c.})\Big],
    \end{aligned} 
\end{equation}
where we notice that the dipole-dipole Hamiltonian is invariant under transformation to the rotating frame of the electric oscillating field~\cite{wil_li_le_2024}.
Here, detuning and Rabi frequency for MW and laser fields are
\begin{equation}
    \begin{aligned}
        \Delta_{\mathrm{MW}} &= E_{P} - E_{S} - \omega_{\mathrm{MW}}, & \Delta_{\mathrm{L}} &= E_{S} - E_{G} - \omega_{\mathrm{L}}; \\
        \Omega_{\mathrm{MW}} &= e\mathcal{E}_{\mathrm{MW}} \bra{S}\!z\!\ket{P}, & \Omega_{\mathrm{L}} &= e\mathcal{E}_{\mathrm{L}} \bra{G}\!z\!\ket{S}.
    \end{aligned}
\end{equation}
To make manifest the MW-dressing, we diagonalize the Hamiltonian describing the Rydberg states manifold,
\begin{equation}
    \begin{aligned}
        \mathcal{H}^{\mathrm{Ryd}}_{\mathrm{in}} &= \sum^{N}_{i = 1} \Big[\Delta_{\mathrm{L}} \ket{S}\!\bra{S}_{i} + (\Delta_{\mathrm{L}} + \Delta_{\mathrm{MW}})\ket{P}\!\bra{P}_{i} \\
        &\qquad \qquad \qquad \qquad \quad- \frac{\Omega_{\mathrm{MW}}}{2}(\ket{S}\!\bra{P}_{i} + \mathrm{H.c.})\Big], \\
        &= \sum^{N}_{i = 1}\big[\Delta_{-} \ket{-}\!\bra{-}_{i} + \Delta_{+} \ket{+}\!\bra{+}_{i}\big].
    \end{aligned} 
\end{equation}
This Hamiltonian generates dressed normalized eigenstates and frequency detunings (i.e., eigenvalues), which are given by
\begin{equation}
    \begin{aligned}
        \ket{\pm} &= \frac{1}{\sqrt{2}}\big[c_{\pm} \ket{P} \pm c_{\mp}\ket{S}\big], \\
        \Delta_{\pm} &= \Delta_{\mathrm{L}} + \frac{ \Delta_{\mathrm{MW}} \pm \sqrt{\Delta^{2}_{\mathrm{MW}} + \Omega^{2}_{\mathrm{MW}}}}{2}.
    \end{aligned}
    \label{eq:detuning}
\end{equation}
The normalization coefficients $c_{\pm}$ are given by
\begin{equation}
    c_{\pm} = \pm \sqrt{1 + \frac{\Delta_{\mathrm{MW}}}{\sqrt{\Delta^{2}_{\mathrm{MW}} + \Omega^{2}_{\mathrm{MW}}}}}.
\end{equation}
For the parameter regime studied here, we consider that the energy splitting of MW-dressed Rydberg states is sufficiently large such that we can neglect the off-resonant coupling of the laser field of frequency $\omega_{\mathrm{L}}$ with the higher-energy dressed Rydberg state $\ket{+}$.
In addition, we tune the MW-field resonantly with the energy difference between Rydberg states $\ket{S}$ and $\ket{P}$, that is, $\omega_{\mathrm{MW}} = E_{P} - E_{S}$.
Under these assumptions, we obtain an effective two-level system consisting of the dressed Rydberg state $\ket{\uparrow} = \ket{-} = \frac{1}{\sqrt{2}}(\ket{P} - \ket{S})$ and low-lying ground state $\ket{\downarrow} = \ket{G}$, as seen in Fig.~\ref{fig:mw_scheme}(b).
The chosen MW-dressed Rydberg state then possess the dipole moments $e\left|\!\bra{\uparrow}\!z\!\ket{\uparrow}\!\right| \propto e\left|\!\bra{S}\! z\! \ket{P}\!\right|$.
To investigate how the magnetic field-induced mix affects the dipole moments of dressed Rydberg states, we show the dipole matrix element $\bra{E_{S}}\!z\!\ket{E_{P}}$ in Fig.~\ref{fig:mw_scheme}(c) as a function of the magnetic field strength $B$.

\begin{figure}
    \centering
    \includegraphics[scale=0.4]{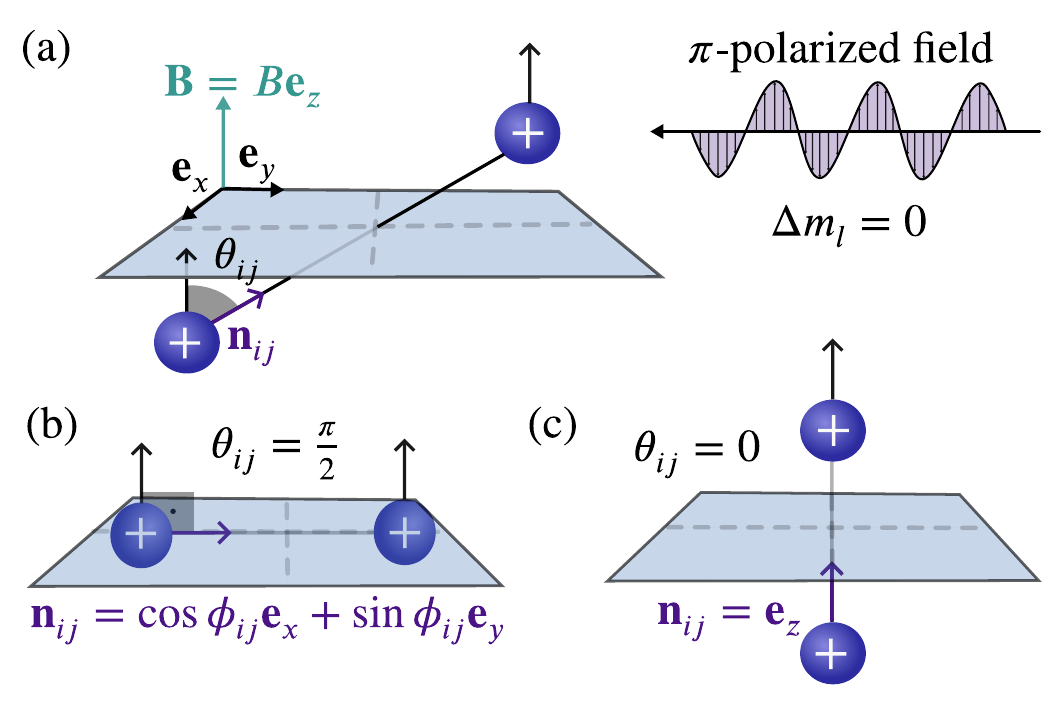}
    \caption{\textbf{Two-body interaction coordinates for linear and planar ion crystals.}
    (a) Two interacting Rydberg ions in a Penning trap.
    The schematic depicts the normalized interparticle axis vector between ions $i$ and $j$, $\mathbf{n}_{ij}$, resulting in the angles $\theta_{ij}$ for the magnetic field $\mathbf{B}$ along the $z$-direction, i.e., $\mathbf{B} = B \mathbf{e}_{z}$; the schematic also highlights that the $\pi$-polarized field along the $z$-axis results in $m_{l}$-conserving transitions.
    (b) Strong axial confinement.
    (c) Strong radial confinement. 
    In the latter, we have defined the azimuth angle $\phi_{ij} = \arctan\big(\tfrac{Y_{i} - Y_{j}}{X_{i} - X_{j}}\big)$.}
    \label{fig:angle}
\end{figure}

To express the internal Hamiltonian as a two-level system with states $\ket{\downarrow}$ and $\ket{\uparrow}$, it is useful to define the angle between the interparticle axis $\mathbf{n}_{ij}$ and the $z$-direction, 
\begin{equation*}
    \theta_{ij} = \cos^{-1}\bigg(\frac{Z_{i} - Z_{j}}{R_{ij}}\bigg),
\end{equation*}
shown in Fig.~\ref{fig:angle}(a).
In the Paschen-Back regime, the internal Hamiltonian reads
\begin{equation}
    \begin{aligned}
        \mathcal{H}_{\mathrm{in}} &= \sum^{N}_{i = 1}  \big(\Delta \mathcal{P}^{\uparrow}_{i} + \Omega \sigma^{x}_{i}\big) \\
        &+ \frac{1}{2}\frac{e^{2}}{4\pi \epsilon_{0}} \sum^{N}_{\substack{i,j=1\\i \ne j}} \frac{\left|\bra{S}z\ket{P}\right|^{2}}{R^{3}_{ij}} \mathcal{P}^{\uparrow}_{i}\mathcal{P}^{\uparrow}_{j} \, \big(1 -3\cos^{2}\!\theta_{ij}\big)
    \end{aligned}
\end{equation}
Here, we have discarded the dipole matrix elements $e \left|\!\bra{\downarrow}\!z\!\ket{\uparrow}\!\right| \propto e\left|\!\bra{G}\!z\!\ket{S}\!\right|$, since the dipole matrix elements associated with Rydberg states are much larger than those of the ground-Rydberg transition, i.e., $e\left|\!\bra{S}\!z\!\ket{P}\!\right| \gg e\left|\!\bra{G}\!z\!\ket{S}\!\right|$.
Additionally, we have defined the projector onto the dressed Rydberg state and the laser drive on the site $i$: $\mathcal{P}^{\uparrow} = \ket{\uparrow} \! \bra{\uparrow}$  and $\sigma^{x} = \ket{\downarrow}\!\bra{\uparrow} + \ket{\uparrow}\!\bra{\downarrow}$; frequency detuning and Rabi frequency are $\Delta = \Delta_{-}$ and $\Omega = \Omega_{\mathrm{L}}/(2\sqrt{2})$, respectively.

The charge-dipole Hamiltonian can be analogously written in terms of the two-level system states.
In the Paschen-Back regime, it becomes
\begin{subequations}
    \begin{align}
        \label{eq:v_dc}
        \mathcal{H}_{\mathrm{cd}} &= \frac{1}{2}\frac{e^{2}}{4\pi \epsilon_{0}} \sum^{N}_{\substack{i,j=1\\i \ne j}} \frac{\bra{S}z\ket{P}}{R^{2}_{ij}}\big(\mathcal{P}^{\uparrow}_{i} - \mathcal{P}^{\uparrow}_{j}\big)\, \cos \theta_{ij}.
    \end{align}
\end{subequations}
In terms of the angles $\theta_{ij}$, it follows directly that 2D crystals confined to the $xy$-plane have vanishing charge-dipole Hamiltonian ($\theta_{ij} = \pi/2$), and consequently there is no internal-external coupling in this situation (see Sec.~\ref{sec:co_int_ext}).

\section{Planar three-ion crystal}
\label{sec:three_ion}

We derive a triangular spin Hamiltonian using a planar crystal with three ions --- the building block of a frustrated quantum magnet.
To this end, we compute the electrostatic interactions among Rydberg states in the three-ion crystal, with particular emphasis on the attainable dipole–dipole interaction strengths.
We find that the internal and external degrees of freedom remain decoupled, while interaction energies on the order of MHz can be achieved.
We further discuss finite-temperature and vibrational effects in the resulting interactions.
Moreover, we present estimates for dephasing and decoherence processes due to fluctuating fields and finite Rydberg state lifetimes.

\subsection{External motion and normal modes of vibration}

We consider an ion crystal strongly confined along the axial direction, forming an equilateral triangular configuration in the $xy$-plane.
For this system, we obtain the equilibrium distances $R^{0}_{12} = R^{0}_{13} = R^{0}_{23} = R_{0}$, with
\begin{equation}
    R_{0} =\bigg(\frac{3e^{2}}{4 \pi\epsilon_{0}M\omega^{2}_{\rho}}\bigg)^{1/3}.
    \label{eq:eq_position}
\end{equation}
The equilibrium inter-ion distance $R_{0}$ depends on the radial confinement frequency $\omega_{\rho}$, which itself is determined by the Penning-trap parameters and the collective rotation frequency of the crystal. 
Consequently, variations of the crystal rotation frequency modify both the equilibrium geometry and the dipole-dipole interaction strength. 
In particular, stronger radial confinement leads to smaller equilibrium separations and therefore enhanced interactions due to the approximate scaling $V_{\mathrm{dd}}\propto R_{0}^{-3}$.
Throughout this work, we consider the standard operating regime in which the crystal rotates at an angular frequency $\omega_{\mathrm{c}}/2$, yielding stable planar configurations. 
More generally, however, the rotation frequency constitutes an additional experimental control parameter that can be used to tune the interaction scale and collective mode spectrum, although excessively weak radial confinement may compromise the stability of the planar crystal.

We expand the external Hamiltonian in terms of small oscillations about the center-of-mass equilibrium positions, $\delta \mathbf{R}_{i} = \mathbf{R}_{i} - \mathbf{R}^{0}_{i}$.
By writing the small oscillations $\delta \mathbf{R}_{i}$ in terms of creation and annihilation operators $a_{\alpha}$ and $a^{\dagger}_{\alpha}$, the Hamiltonian describing the vibrational external dynamics reads
\begin{equation}
    \mathcal{H}_{\mathrm{ex}} = \sum^{5}_{\alpha = 1} \omega_{\alpha}\big(a^{\dagger}_{\alpha}a_{\alpha} + 1/2\big).
\end{equation}
The normal-mode frequencies $\omega_{\alpha}$ are provided by the radial frequency $\omega_{\rho}$ multiplied by mode-dependent scaling factors.
The confinement of the external motion to the $xy$ plane yields six degrees of freedom and hence six normal modes. 
One of these, however, is a soft mode with vanishing frequency $\omega_{0}$, which corresponds to in-plane rotations.
A minor anisotropy in the radial confinement (assuming that $\omega_{\rho}$ can be rewritten as $\omega_{x}$ and $\omega_{y}$ for $x$- and $y$-axis, with $\omega_{x} \ne \omega_{y}$) breaks the rotational symmetry, and gives that mode a finite frequency.
The remaining frequencies and their respective normal modes are $\omega_{1} = \omega_{2} = \omega_{\rho}$, the center of mass modes, $\omega_{3} = \omega_{4} = \sqrt{3/2}\omega_{\rho}$, the rocking modes, and $\omega_{5} = \sqrt{3}\omega_{\rho}$, the breathing modes. 
The derivation of the normal modes for the 2D three-ion crystal is detailed in App.~\ref{app:normal_modes}.

\subsection{Spin-spin interaction strength and internal-external decoupling}

\begin{figure}
    \centering
    \includegraphics[scale=0.38]{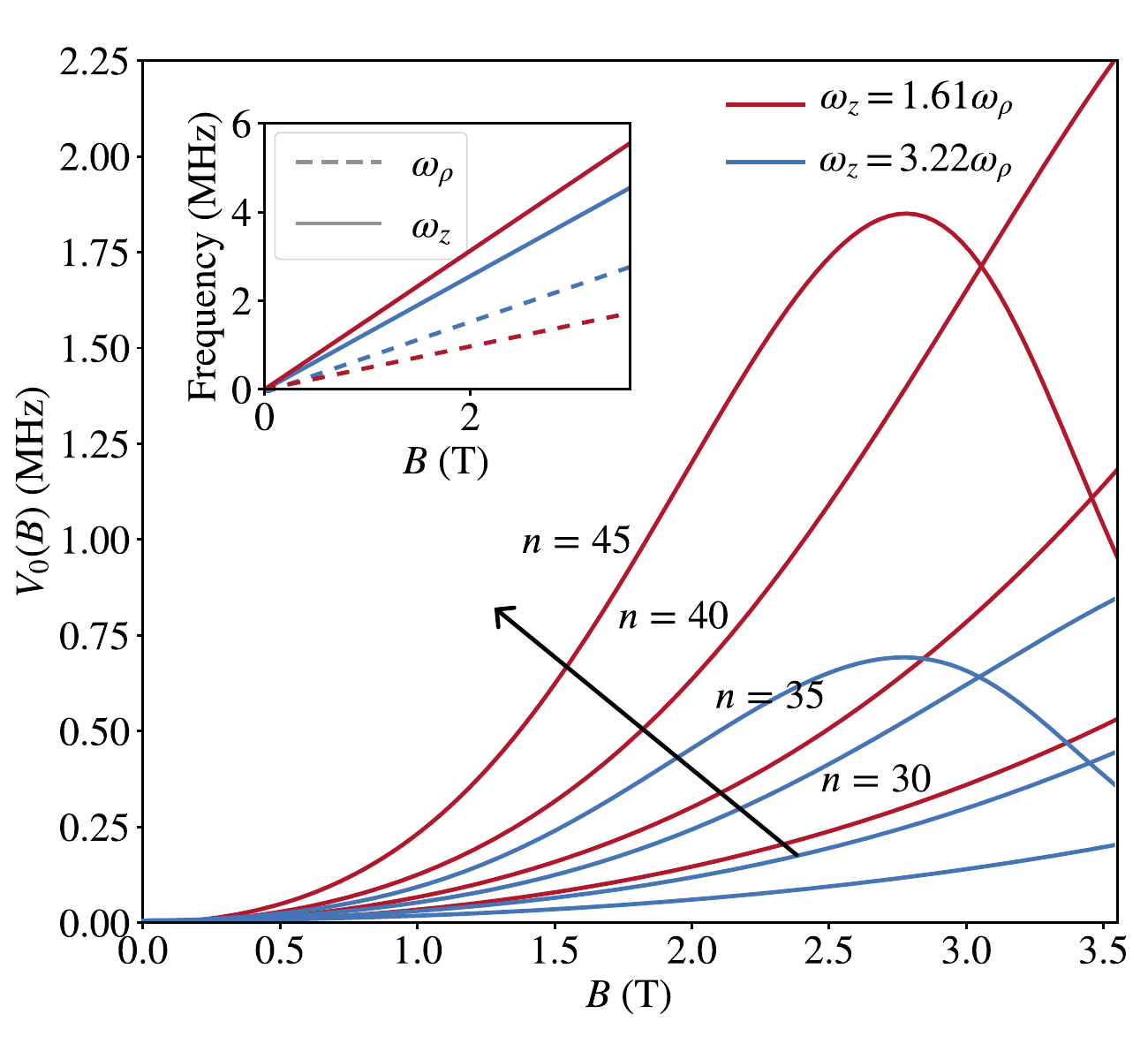}
    \caption{\textbf{Dipole-dipole interaction strength in a planar three-ion crystal.}
    The figure shows the dipole-dipole interaction strength versus magnetic field strength $B$, for different values of the principal quantum number $n$.
    The ratio between axial and radial frequencies is set $\omega_{z}/\omega_{\rho} = 1.61$ (red) and $\omega_{z}/\omega_{\rho} = 3.22$ (blue), respectively.
    The inset shows axial (dashed lines) and radial (solid lines) frequencies as a function of the magnetic field strength $B$.
    The principal quantum numbers are $n = 30, 35, 40$ and $45$.
    Values here are obtained with the eigenfunctions for $^{40}\text{Ca}^{+}$ ions.
    }
    \label{fig:panel_results}
\end{figure}

We now turn to the internal electronic dynamics.
For ions located in the $xy$-plane, the angles between the interparticle $\mathbf{n}_{ij}$ and the $z$-direction for each interacting pair of ions are $\theta_{ij}= \pi/2$, see Fig.~\ref{fig:angle}.
The resulting internal Hamiltonian then becomes
\begin{equation}
    \mathcal{H}_{\mathrm{in}} = \Omega\sum^{3}_{i = 1} \sigma^{x}_{i} - \Delta \sum^{3}_{i = 1} \mathcal{P}^{\uparrow}_{i} + \mathcal{H}_{\mathrm{dd}},
\end{equation}
where the charge-dipole coupling vanishes and the dipole-dipole interaction Hamiltonian reads
\begin{equation}
    \mathcal{H}_{\mathrm{dd}} =  \frac{1}{2}\sum^{3}_{\substack{i,j=1\\i \ne j}}V_{\mathrm{dd}}(R_{ij}) \mathcal{P}^{\uparrow}_{i} \mathcal{P}^{\uparrow}_{j}.
\end{equation}
Here, we have defined the distance-dependent dipole-dipole interaction potential, which is given by
\begin{equation}
    V_{\mathrm{dd}} (R_{ij}) = \frac{e^{2}}{4\pi\epsilon_{0} R^{3}_{ij}} \left|\!\bra{S}\!z\!\ket{P}\!\right|^{2},
    \label{eq:dip_dip_int}
\end{equation}
in the Paschen-Back regime.
The force associated with this interaction can lead to spin-phonon coupling.
The coupling strength can be estimated by performing the harmonic expansion of Eq.~\eqref{eq:dip_dip_int} to leading-order contributions in the spin-phonon coupling.
Considering the linear phonon term, the harmonic expansion generates the following potential, given in terms of the operators $a_{\alpha}$ and $a^{\dagger}_{\alpha}$:
\begin{equation}
    \begin{aligned}
        V_{\mathrm{dd}} (R_{ij}) \approx V_{0} + \sum^{5}_{\alpha = 1} \mathcal{W}^{\alpha}_{ij}(a^{\dagger}_{\alpha} + a_{\alpha}).
    \end{aligned}
    \label{eq:dip_exp}
\end{equation}
Here, we have defined the dipole-dipole interaction strength $V_{0} = V_{\mathrm{dd}}(R_{0})$.
The coefficients $\mathcal{W}^{\alpha}_{ij}$ constitute normal mode-dependent coupling strengths.
In particular, they vanish for the center-of-mass modes, i.e., $\mathcal{W}^{1}_{ij} = \mathcal{W}^{2}_{ij} = 0$, since these modes only act on the overall motion of the three-ion crystal.
The additional coefficients are given by 
\begin{equation*}
    \begin{aligned}
        -2\mathcal{W}^{3}_{12} = \mathcal{W}^{3}_{13} = -2\mathcal{W}^{3}_{23} &= 2\ell_{3}V_{0}/R_{0}, \\
        -\mathcal{W}^{4}_{12} = 2\mathcal{W}^{4}_{13} = 2\mathcal{W}^{4}_{23} &= 2\ell_{4}V_{0}/R_{0}, \\
        \mathcal{W}^{5}_{12} = \mathcal{W}^{5}_{13} = \mathcal{W}^{5}_{23} &= 2\ell_{5}V_{0}/R_{0},
    \end{aligned}   
\end{equation*}
where $\ell_{\alpha} = \sqrt{1/(2M\omega_{\alpha})}$ the characteristic length of each normal mode.
Within the parameter range of Fig.~\ref{fig:panel_results}, the gradient terms in Eq.~\eqref{eq:dip_exp} that couple spin and vibrational motion are negligible, with $\mathcal{W}^{\alpha}_{ij}/V_{0} \sim 10^{-3}$.
Therefore, the internal and external dynamics are effectively decoupled, justifying a description based solely on the spin degrees of freedom of the internal electronic states.
In practice, a negligible spin-phonon coupling also enables high-fidelity spin-state readout in experiments.

To study the achievable spin–spin coupling, we calculate the dipole–dipole interaction strength arising from magnetic-field–dressed states,
\begin{equation}
    V_{0}(B) = \frac{e^{2}}{4\pi\epsilon_{0} R_{0}^{3}}
\left|\!\bra{E_{S}} z \ket{E_{P}}\!\right|^{2}(B),
\end{equation}
where the states $\ket{E_{S}} = \ket{E_{S}(B)}$ and $\ket{E_{P}} = \ket{E_{P}(B)}$ are defined in Eq.~\eqref{eq:ad_states}.
This expression also describes the interaction strength in the $l$-mixing regime, where the diamagnetic term becomes important.
Since $V_{0} \propto M^{-2}$, different and less massive ion species may be utilized to generate stronger interactions, e.g., $\text{Be}^{+}$ or  $\text{Mg}^{+}$, see App.~\ref{app:ion_species}.
In the case of the $^{9}\text{Be}^{+}$ isotope, the non-vanishing nuclear spin also couples with the trap magnetic field, creating hyperfine Zeeman splitting.
Furthermore, since
\begin{equation*}
    V_{0} \propto \omega_{\rho}^{2}
    = \frac{\omega_{\mathrm{c}}^{2} - 2\omega_{z}^{2}}{4},
\end{equation*}
there is a trade-off between maximizing the dipole–dipole interaction strength and maintaining the strong axial confinement required for planar crystals.
For a three-ion crystal, planar confinement is achieved when \textcolor{blue}{$\omega_{z} \gtrsim 1.61\,\omega_{\rho}$} ~\cite{schiffer1993, dubin1993}.
Imposing this condition, Fig.~\ref{fig:panel_results} shows $V_{0}$ for several principal quantum numbers $n$ at fixed relative confinement strength $\omega_{z}/\omega_{\rho}$, with the inset displaying the corresponding confinement frequencies $\omega_{z}$ and $\omega_{\rho}$.

The curves in Fig.~\ref{fig:panel_results} further show that $V_{0}$ decreases as the diamagnetic term becomes relevant, reflecting the induced mixing between different orbital angular momentum states.
Since the diamagnetic term scales as $n^{4}$, this effect is more pronounced for higher $n$.
Thus, while larger $n$ enhances the dipole moment, it also restricts the magnetic-field range over which $V_{0}$ remains strong.
For $\omega_{z} = 1.61\omega_{\rho}$, we obtain dipole–dipole interaction strengths on the MHz scale.
For comparison, frustrated Ising interactions in a three-ion crystal were realized in Ref.~\cite{kim2010} with coupling strengths of approximately $2\,\text{kHz}$.
In contrast, the interactions considered here are approximately three orders of magnitude stronger.
This regime enables the observation of frustration and entanglement in a two-dimensional quantum magnet formed by a Rydberg-excited three-ion crystal, well within experimentally relevant coherence times~\cite{langer2005, wang2017, wang2021}.

\subsection{Cooling and finite temperature effects}

In the following, we discuss cooling strategies and finite temperature effects in the quantum simulator.
Given its dependency on the ion's motion, Eq.~\eqref{eq:dip_exp} indicates that thermal occupation of the vibrational modes induces fluctuations in the dipole-dipole interactions.
In particular, low-frequency magnetron-dominated modes (commonly referred to as $\mathbf{E}\times\mathbf{B}$ modes), whose frequencies depend on $\omega_{-} \approx \omega_{\mathrm{c}}/2 - \omega_{\rho}$, represent the weakest confined collective degrees of freedom in large Penning-trap crystals and are therefore the most susceptible to residual thermal excitation~\cite{wang2013, shan_tan_2020}.
In the present platform, however, the effective spin dynamics originate from direct dipole-dipole interactions rather than phonon-mediated couplings, substantially reducing the sensitivity to finite motional occupation.
Consequently, thermally induced fluctuations remain perturbatively small even for moderately occupied collective modes.
The primary experimental requirement is therefore the preservation of crystal stability and inter-ion spacing, rather than near-ground-state cooling of all radial modes.
Established Penning-trap techniques, including axialization~\cite{powell_2002}, rotating-wall confinement~\cite{to_bri_2016} and mode-coupling-assisted cooling~\cite{gorman_2014, johnson_2024, hou_2024, johnson_2025}, provide effective mechanisms for stabilizing these low-frequency degrees of freedom over experimentally relevant timescales.

To quantify these effects, we note that thermal occupation enhances the fluctuation amplitude according to
\begin{equation}
    \delta R_{\alpha} = \ell_{\alpha }\sqrt{\big\langle (a^{\dagger}_{\alpha} + a_{\alpha})^{2} \big\rangle}
    = \ell_{\alpha } \sqrt{2\bar{n}_{\alpha} + 1},
\end{equation}
for each vibrational mode $\alpha$.
Residual motion then produces relative interaction fluctuations of the order
\begin{equation}
    \frac{\delta V_{0}}{V_{0}}
    \sim
    10^{-3}\sqrt{2\bar{n}_{\alpha} + 1}.
\end{equation}
The Bose-Einstein distribution gives the phonon occupation number, $\bar{n}_{\alpha} = \left( \e^{\omega_{\alpha}/k_{\mathrm{B}}T} -1 \right)^{-1}$.
A relative instability below $1\,\%$ of the spin-spin couplings is therefore maintained provided $\bar{n}_{\alpha} \lesssim 50$ for the most strongly coupled modes, while less stringent accuracy requirements permit substantially larger occupations.
For a representative mode frequency of $\omega_{\alpha} = 2\pi \times 200 \, \mathrm{kHz}$, this occupation corresponds to a temperature of approximately $T_{\alpha} \approx 0.5 \, \mathrm{mK}$.
Such temperatures are achievable at the Doppler-cooling limit for $^{40}\mathrm{Ca}^{+}$ ions~\cite{roos_2000}.

\subsection{Microwave inhomogeneity and field-fluctuation effects}
\label{sec:MWI_fluctuation}

In the following, we discuss the effects of imperfections arising from MW inhomogeneities and field fluctuations and strategies to mitigate them.
In principle, the MW fields may exhibit undesired spatial variations in intensity and polarization impurity.
The corresponding wavelengths are $\lambda_{\mathrm{MW}} = 2\pi c/\omega_{\mathrm{MW}} \approx 3 \, \mathrm{mm}$.
By contrast, the size of trapped-ion crystals typically remains on the hundred-micrometer scale even for large systems.
Fig.~\ref{fig:shift_grad}(a) shows that a crystal with $N = 40$ ions reaches a maximum inter-ion distance of $R^{\mathrm{max}}_{0} < 100 \, \mu\mathrm{m}$.
Consequently, the ratio satisfies $R^{\mathrm{max}}_{0}/\lambda_{\mathrm{MW}} \lesssim 0.1$, substantially suppressing the effects of spatial MW intensity modulations.

Experimentally, it has been observed that a MW field with well-defined amplitude and polarization outside the ultra-high-vacuum chamber can arrive at the center of the ion trap with both quantities significantly distorted~\cite{Bao_2025}. 
Such polarization impurities may introduce errors by coupling to Rydberg states with magnetic quantum numbers $m_l$ different from those of the target state. 
One approach to mitigate this effect is the use of micro-structured traps with integrated antenna structures that provide enhanced control over the polarization and propagation direction of the injected microwave field. 
An example of this technology, albeit operating at substantially lower frequencies, is the integrated microwave enhancement resonator developed for an electron Paul trap~\cite{mikhailovskii_2026}.
In addition, the operating regime considered here benefits from a strong Zeeman splitting, which lifts the degeneracy between states with different $m_l$ by several hundred GHz. Consequently, state selectivity can be achieved not only through polarization control but also through frequency discrimination. 
This large spectral separation substantially suppresses unwanted off-resonant couplings and reduces the sensitivity of the scheme to residual polarization imperfections and fluctuations.

Fluctuations of the magnetic and microwave fields may constitute an important source of qubit dephasing. 
In a two-ion Rydberg gate experiment, MW power instability was identified as the dominant contribution to the gate error budget~\cite{zhang_po_wei_2020}. 
Motivated by these observations, in App.~\ref{app:fluc_fields} we derive the dephasing rates associated with magnetic- and microwave-field fluctuations and estimate their magnitude using experimentally reported noise levels. 
Based on these results, we quantitatively evaluate and compare the impact of the relevant decoherence mechanisms in the system.

\subsection{Decay rates from different dephasing and decoherence sources}

\textcolor{blue}{
\begin{table}
\caption{\textbf{Estimated decay rates and dephasing strengths for different channels.}
The decay rates are calculated using the AlkCalc library~\cite{alkcalc_2026}.
The dephasing strengths are obtained from root-mean-square field noise amplitudes found in recent experiments~\cite{zhang_po_wei_2020, mcmahon_2022}; the calculations are described in App.~\ref{app:fluc_fields}.}
\begin{tabular}{ |c| c| } 
   \hline 
   Source & Contribution  \\
  \hline
  Natural state lifetime & $\Gamma_{\uparrow} = 2\pi \times 15.0 \,\mathrm{kHz}$  \\ 
  \hline
  Blackbody radiation  ($T = 300\,\mathrm{K}$) & $\Gamma_{\mathrm{BBR}} = 2\pi \times 6.6 \,\mathrm{kHz}$  \\ 
  \hline
  Fluctuating magnetic field & $\Gamma_{B} = 2\pi \times 1.1 \, \mathrm{kHz}$ \\ 
  \hline
  Fluctuating MW-field & $\Gamma_{\mathrm{MW}} = 2\pi \times 3.0 \, \mathrm{kHz}$\\
  \hline 
\end{tabular}
\label{tab:decay}
\end{table}
}

\label{sec:decay}
We now summarize all relevant dephasing and relaxation mechanisms and compare them with the achievable dipole-dipole interaction strengths.
We begin by considering the finite lifetime of the dressed Rydberg state $\ket{\uparrow}$.
Since $\ket{\uparrow}$ is an equal-weight superposition of the states $\ket{S}$ and $\ket{P}$, its population decays through the decay channels associated with both constituents.
The population of the dressed state evolves as
\begin{equation*}
    P_{\uparrow}(t)=\frac{1}{2}\big(e^{-t/\tau_{P}} +e ^{-t/\tau_{S}}\big).
\end{equation*}
An effective decay rate, obtained from the initial population decay, is
\begin{equation*}
    \Gamma_{\uparrow}\approx\frac{\tau^{-1}_{S}+\tau^{-1}_{P}}{2}.
\end{equation*}
For an example with a principal quantum number $n=45$, the natural lifetimes of the Rydberg states yield
$\Gamma_{\uparrow} = 2\pi \times 15.0\,\mathrm{kHz}$.
Additional decay induced by blackbody radiation leads to a contribution $\Gamma_{\mathrm{BBR}} = 2\pi\times6.6\,\mathrm{kHz}$ at room temperature ($T = 300\,\mathrm{K}$).
In a cryogenic environment, where the temperature can be approximated as $T \lesssim 5\,\mathrm{K}$, this contribution becomes negligible.
A summary of the decay rates and dephasing strengths arising from finite Rydberg-state lifetimes, as well as from fluctuating magnetic and MW fields, is presented in Tab.~\ref{tab:decay}.
Even after accounting for all decoherence channels, the finite lifetime of the Rydberg state remains the dominant limitation.
Importantly, the achievable dipole-dipole interaction strengths are still significantly larger than the effective decay rate of the dressed Rydberg state, i.e., $V_0 \gg \Gamma_{\uparrow}$, allowing for coherent interaction dynamics on timescales much shorter than those associated with decoherence.

\subsection{Spin Hamiltonian and ground state physics}

In the following, we examine the ground state of the internal Hamiltonian of the three-ion crystal,
\begin{equation}
        \mathcal{H}_{\mathrm{in}} = \Omega \sum^{3}_{i = 1} \sigma^{x}_{i} - \Delta \sum^{3}_{i = 1} \mathcal{P}^{\uparrow}_{i} + \frac{V_{0}}{2} \sum^{3}_{\substack{i,j=1\\i \ne j}} \mathcal{P}^{\uparrow}_{i} \mathcal{P}^{\uparrow}_{j},
    \label{eq:spin_ham}
\end{equation}
akin to an Ising model, where $V_{0}$ represents the spin-spin coupling.
This Hamiltonian can be mapped into a geometrically frustrated spin system.
In this scenario, pairwise interaction energies cannot be simultaneously minimized, leading to a sixfold degenerate ground state when the Rabi frequency vanishes, $\Omega = 0$~\cite{le_ji_ga_2015, chan_shen_2022}.
This degeneracy is lifted for non-vanishing Rabi frequency $\Omega$, resulting in the selection of entangled states~\cite{mo_son_2000, moess_son_2001, gui_huan_2023}.

\begin{figure}
    \centering
  \includegraphics[scale=0.4]{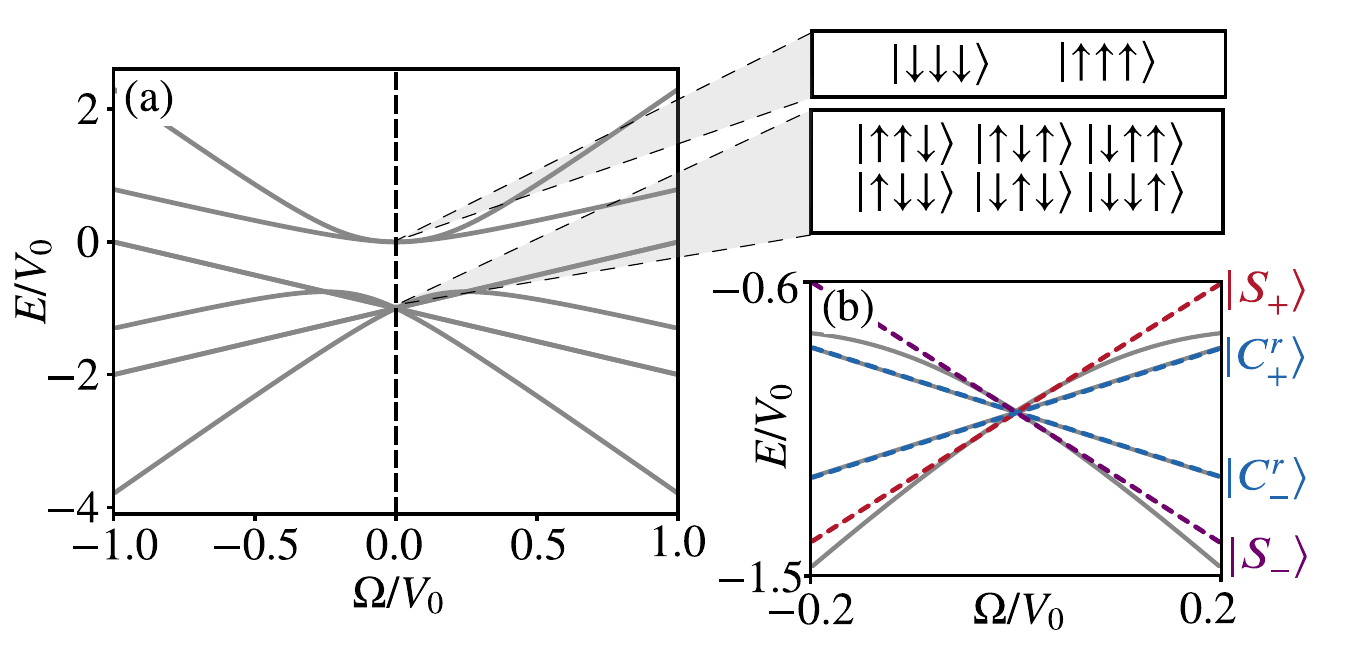}
    \caption{\textbf{Spectrum of the spin Hamiltonian.}
    (a) Exact spectrum of the Hamiltonian $\mathcal{H}_{\mathrm{in}} = \mathcal{H}_{0} + \delta\mathcal{H}$ obtained by numerical diagonalization under the facilitation condition $V_{0} = -\Delta$.
    At zero Rabi frequency ($\Omega = 0$, vertical dashed line), the ground-state manifold is sixfold degenerate; the corresponding frustrated spin configurations are shown in the inset.
    (b) For finite Rabi frequency ($\Omega \neq 0$), the transverse-field term $\delta\mathcal{H}$ lifts this degeneracy and produces entangled eigenstates within the formerly degenerate manifold.
    The dashed lines indicate the first-order perturbative energy corrections associated with the eigenstates illustrated on the right.
    }
    \label{fig:gstate}
\end{figure}

We briefly recall the spectrum of the spin Hamiltonian, which we conveniently write as $\mathcal{H}_{\mathrm{in}} = \mathcal{H}_{0} + \delta \mathcal{H}$~\cite{po_shi_2016}.
We start by considering the state space of the Hamiltonian
\begin{equation}
    \mathcal{H}_{0} = - \Delta \sum^{3}_{i = 1} \mathcal{P}^{\uparrow}_{i} + \frac{V_{0}}{2} \sum^{3}_{\substack{i,j=1\\i \ne j}} \mathcal{P}^{\uparrow}_{i} \mathcal{P}^{\uparrow}_{j}.
\end{equation}
The Hamiltonian $\mathcal{H}_{0}$ is diagonal on the basis of pure product states $\ket{\bm{\sigma}} = \ket{\sigma_{1} \sigma_{2} \sigma_{3}}$, with $\ket{\sigma} = \ket{\downarrow}, \ket{\uparrow}$.
By tuning the laser frequency to $\Delta + V_{0} = 0$, the laser becomes resonant with transitions whose energy cost is offset by the dipolar interaction.
In a triangular geometry, this condition does not single out a unique excitation pattern, but instead brings several competing many-body configurations close in energy, including states with different numbers of Rydberg excitations.
The resulting \textit{facilitation condition} therefore imposes a kinetic constraint on the dynamics, promoting frustration by making the allowed transitions explicitly dependent on the many-body configuration~\cite{weimer2008, lesanovsky2012}.
In this regime, the spectrum separates into two parts: the energy $E = 0$ contains the states $\ket{\downarrow \downarrow \downarrow}$ and $\ket{\uparrow \uparrow \uparrow}$ and the ground-state energy $E = - \Delta$ contains the degenerate set formed by the states
\begin{equation*}
    \ket{\uparrow \downarrow \downarrow}, \ket{\downarrow \uparrow \downarrow}, \ket{\downarrow \downarrow\uparrow}, \ket{\uparrow \uparrow \downarrow}, \ket{\uparrow \downarrow \uparrow}, \: \text{and} \: \ket{\downarrow \uparrow\uparrow}.
\end{equation*}
We proceed by introducing  $\delta \mathcal{H} = \Omega \sum_{i} \sigma^{x}_{i}$ perturbatively.
Degenerate perturbation theory in this case is simplified by adopting a basis that transforms according to the irreducible representations of the symmetry group of the equilateral triangle, generated by its rotations and reflections~\cite{step_1964, sch_wal_1977, xi_zhao_2024}.
This means that the basis states are chosen such that they remain within invariant subspaces of the group action~\cite{lhuillier2002frustrated}.
Such a representation generates the pair of symmetric states given by
\begin{equation*}
    \begin{aligned}
        \ket{S_{1}} &= \frac{1}{\sqrt{3}} \big[\ket{\uparrow \downarrow \downarrow} + \ket{\downarrow \uparrow \downarrow} + \ket{\downarrow \downarrow\uparrow}\big], \\
        \ket{S_{2}} &= \frac{1}{\sqrt{3}} \big[\ket{\uparrow \uparrow \downarrow} + \ket{\uparrow \downarrow \uparrow} + \ket{\downarrow \uparrow\uparrow}\big],
    \end{aligned}
\end{equation*}
and the four states in the form
\begin{equation*}
    \begin{aligned}
        \ket{C^{r}_{1}} &= \frac{1}{\sqrt{3}} \big[\ket{\uparrow \downarrow \downarrow} + \e^{\frac{2\pi \mathrm{i}r}{3}}\ket{\downarrow \uparrow \downarrow} + \e^{\frac{4\pi \mathrm{i}r}{3}}\ket{\downarrow \downarrow\uparrow}\big], \\
        \ket{C^{r}_{2}} &= \frac{1}{\sqrt{3}} \big[\ket{\uparrow \uparrow \downarrow} + \e^{\frac{2\pi \mathrm{i}r}{3}}\ket{\uparrow \downarrow \uparrow} + \e^{\frac{4\pi \mathrm{i}r}{3}}\ket{\downarrow \uparrow\uparrow}\big],
    \end{aligned}
\end{equation*}
where $r = \pm 1$~\cite{moe_son_chan_2000, po_co_2013}.
We consider first-order corrections to the eigenstates and eigenvalues of $\mathcal{H}_{\mathrm{in}}$, obtaining states and energies 
\begin{equation}
    \begin{aligned}
    \ket{S_{\pm}} = \frac{1}{\sqrt{2}} (\ket{S_{1}} \pm \ket{S_{2}}), \; \text{with}\quad  E^{(1)} = -(\Delta \pm 2 \Omega); \\
    \ket{C^{r}_{\pm}} = \frac{1}{\sqrt{2}} (\ket{C^{r}_{1}} \pm \ket{C^{r}_{2}}), \; \text{with} \quad E^{(1)} = -(\Delta \pm \Omega).
    \end{aligned}
\end{equation}
Here, the sixfold degeneracy splits into four distinct energy levels, with the ground-state energy $E_{\mathrm{GS}} = - (\Delta + 2\Omega)$. 
Accordingly, the state $\ket{S_{\pm}}$ (depending on the sign of $\Omega$) is selected under the perturbation introduced by $\delta \mathcal{H}$~\cite{da_1994, jan_ma_2024}.  
Fig.~\ref{fig:gstate}(a) shows the spectrum of $\mathcal{H}_{\mathrm{in}}$ for $V_{0} = - \Delta$, while Fig.~\ref{fig:gstate}(b) displays the first-order perturbative energies with their associated eigenstates.
To ensure that the Rabi coupling only lifts degeneracies within the frustrated manifold, without qualitatively modifying the underlying interaction-dominated physics, we require $\Omega/V_{0} \ll 1$.

Having established the structure of the energy spectrum and the nature of the ground state, we now connect these findings to experimentally accessible conditions.
The ground-state structure described above can be obtained for Rydberg ions in a Penning trap by setting $B = 1.75\,\text{T}$ and $\beta = 0.51 \times 10^{6}\,\text{V/m}^{2}$, which results in confinement frequencies $\omega_{z} = 2\pi \times 350\,\text{kHz}$ and $\omega_{\rho} = 2\pi \times 220\,\text{kHz}$ and a spin–spin coupling strength $V_{0} \approx 1\,\text{MHz}$ for $n = 45$.
This parameter regime is consistent with the current experimental capabilities of Penning-trap architectures, as well as Rydberg-excited ions~\cite{ma_go_jo_2013, goo_stu_2016, zhang_po_wei_2020}.
The case example of frustration using a three-ion crystal in a triangular configuration may be extended to a large planar crystal consisting of many such triangular unit cells. 
We expect that this will give rise to a rich physics of frustrated states with interesting multiparticle entanglement properties.

\section{Summary and perspectives}
\label{sec:conclusions}

We have proposed and analyzed a platform for 2D quantum simulation that is based on trapped Rydberg ions in a Penning trap.
Starting from a microscopic model, we derived the single-ion Hamiltonian under external trapping fields and identified the conditions under which Rydberg states remain stable in the Paschen–Back regime. 
Through MW-dressing, these states acquire sizable and tunable electric dipole moments, enabling the implementation of interacting spin systems. 
Our analysis shows that the resulting dipole-dipole interactions reach MHz strengths, enabling the buildup of many-body correlations on timescales much shorter than the intrinsic coherence times of trapped ions.
Finally, we investigated a planar three-ion crystal, demonstrating that such a system leads to a quantum spin model with frustrated interactions and entangled ground states. 

Quantum simulation with trapped Rydberg ions in a Penning trap offers many opportunities: 
in large ion crystals, spatially resolved Rydberg excitation permits the realization of various effective lattice geometries, including triangular, kagome, and hexagonal arrangements~\cite{richerme_2016}, which allow for direct exploration of frustrated magnetism and exotic quantum phases~\cite{Pyka_2013, gla_dal_2015, shan_yu_2022, yan_sa_2022, giu_lu_2022, po_rey_2022, kor_sa_2023, mc_brown_2024, mo_bre_2025}.
Moreover, coupled electron-phonon (vibronic) dynamics can be realized by engineering interactions between Rydberg states and collective modes of the ion crystal. This is opening avenues for the quantum simulation of molecular phenomena such as the Jahn–Teller effect~\cite{po_iva_sch_2012, hem_ma_2018, ma_jo_le_2023, eu_le_2025}.
Furthermore, the extraordinary stability of trapped ion systems together with strong dipolar interactions among Rydberg states allow the study of slow relaxation caused by so-called kinetic constraints. This may open a new window for the investigation of intricate non-equilibrium behavior such as glassy dynamics and trajectory phase transitions~\cite{ol_le_ga_2012, le_ga_2013, pe_le_ga_2018, ga_2018, os_mar_2019}.

\textbf{Data access statement:} The numerical data supporting the findings of this article and the code used to generate them are available on Zenodo~\cite{wmartins2025}.

\begin{acknowledgments}
We are grateful to J.W.P. Wilkinson, S. Euchner, A. Cabot, P.J. Paulino, and R. Côté for valuable discussions.
We acknowledge funding from the Deutsche Forschungsgemeinschaft (DFG, German Research Foundation) through the Research Unit FOR 5413/1, Grant No.~465199066, and through JST-DFG 2024: Japanese-German Joint Call for Proposals on “Quantum Technologies” (Japan-JST-DFG-ASPIRE 2024) under DFG Grant No. 554561799.
We acknowledge support from the U.S. National Science Foundation (NSF) Grants No. PHY-2513221 and No. OSI-2328774.
This work was supported by the QuantERA II programme (project CoQuaDis, DFG Grant No. 532763411) that has received funding from the EU H2020 research and innovation programme under GA No. 101017733. This work was also supported by the ERC grant OPEN-2QS (Grant No. 101164443).

\end{acknowledgments}

\bibliography{main} 

\newpage

\appendix

\begin{widetext}
\section{Derivation of internal, external, and coupling Hamiltonians}
\label{app:derivation}

In this appendix, we derive the single-ion Hamiltonian in the center-of-mass reference frame.
We begin with the single-ion Hamiltonian in terms of center-of-mass and relative coordinates; see Eq.~\eqref{eq:CM_coord},
\begin{equation}
    \begin{aligned}
        H &= \frac{1}{2m_\mathrm{e}} \bigg[\mathbf{p}+ \frac{e m_\mathrm{c}\big(\mathbf{B} \times \mathbf{R}\big)}{2(m_\mathrm{c} + m_\mathrm{e})} + \frac{m_\mathrm{e} \mathbf{P}}{m_\mathrm{c} + m_\mathrm{e}} + \frac{e}{2} \mathbf{B} \times  \mathbf{r}\bigg]^2 +  \frac{1}{2m_\mathrm{c}} \left[\mathbf{p}- \frac{e m_\mathrm{e}\big(\mathbf{B} \times \mathbf{R}\big)}{m_\mathrm{c} + m_\mathrm{e}}  - \frac{m_\mathrm{c}}{m_\mathrm{c} + m_\mathrm{e}}  \mathbf{P}+ e \mathbf{B} \times  \mathbf{r}\right]^2 \\
        &- \boldsymbol{\mu}_{\mathrm{e}}\cdot \mathbf{B}-  e \boldsymbol{\mathcal{E}}(\mathbf{R})\cdot \mathbf{R} + e\bigg(\frac{m^{2}_{\mathrm{c}}}{M^{2}} - \frac{2 m^{2}_{\mathrm{e}}}{M^{2}}\bigg) \boldsymbol{\mathcal{E}}(\mathbf{r})\cdot \mathbf{r} + e\bigg(\frac{m_{\mathrm{c}}}{M} + \frac{2 m_{\mathrm{e}}}{M}\bigg) [\boldsymbol{\mathcal{E}}(\mathbf{r})\cdot \mathbf{R} + \boldsymbol{\mathcal{E}}(\mathbf{r})\cdot \mathbf{R}].
    \end{aligned}
\end{equation}
We proceed with the gauge-invariant transformation $H \mapsto U^{\dagger} H U$, where $U (\mathbf{r}, \mathbf{R}) = \exp\big[\!-\!\mathrm{i} \tfrac{e(m_{\mathrm{c}} + 2m_{\mathrm{e}})}{M}  \mathbf{A}(\mathbf{R}) \cdot \mathbf{r}\big]$~\cite{sch_ce_1991}.
The momentum operators transform as
\begin{equation}
    \begin{aligned}
    U^\dagger \mathbf{p} U &= \mathbf{p} - \frac{e(m_{\mathrm{c}} + 2m_{\mathrm{e}})}{2M} (\mathbf{B}\times \mathbf{R}), \\
    U^\dagger \mathbf{P} U &=  \mathbf{P}+ \frac{e(m_{\mathrm{c}} + 2m_{\mathrm{e}})}{2M} (\mathbf{B}\times \mathbf{r}),
    \end{aligned}
\end{equation}
while the position operators $\mathbf{r}$ and $\mathbf{R}$ remain invariant.
These relations follow from the identity $(\mathbf{r}\times \mathbf{B}) \cdot \mathbf{R} = - (\mathbf{B} \times \mathbf{r}) \cdot \mathbf{R}$, and from the Baker-Campbell-Hausdorff formula. 
For ease of notation, we make $ \gamma  = (m_{\mathrm{c}} + 2m_{\mathrm{e}})/M$.
The single-ion Hamiltonian comes as
\begin{equation}
    \begin{aligned}
    H &= \frac{1}{2m_{\mathrm{e}}} \bigg[\mathbf{p} + \frac{e}{2} \frac{m_{\mathrm{c}} +  \gamma  m_{\mathrm{e}}}{M} (\mathbf{B} \times \mathbf{r})\bigg]^{2} + \frac{1}{2m_{\mathrm{c}}} \bigg[\mathbf{p} - \frac{e}{2} \frac{2m_{\mathrm{e}} +  \gamma  m_{\mathrm{c}}}{M} (\mathbf{B} \times \mathbf{r})\bigg]^{2} + \frac{m_{\mathrm{e}}}{2M^{2}} \bigg[\mathbf{P} - \frac{e}{2} (\mathbf{B} \times \mathbf{R})\bigg]^{2}  \\
    & + \frac{m_{\mathrm{c}}}{2M^{2}} \bigg[\mathbf{P} - \frac{e}{2} \frac{m_{\mathrm{c}}}{M}(\mathbf{B} \times \mathbf{R})\bigg]^{2}+ \frac{1}{M} \bigg[\mathbf{p} + \frac{e}{2} \frac{m_{\mathrm{c}} +  \gamma  m_{\mathrm{e}}}{M} (\mathbf{B} \times \mathbf{r})\bigg] \bigg[\mathbf{P} - \frac{e}{2} (\mathbf{B} \times \mathbf{R})\bigg]  \\
    &- \frac{1}{M} \bigg[\mathbf{p} - \frac{e}{2} \frac{2 m_{\mathrm{e}} +  \gamma m_{\mathrm{c}}}{M} (\mathbf{B} \times \mathbf{r})\bigg] \bigg[\mathbf{P} - \frac{e}{2} \frac{m_{\mathrm{c}}}{M}(\mathbf{B} \times \mathbf{R})\bigg]  - \boldsymbol{\mu}_{\mathrm{e}}\cdot \mathbf{B} -  e \boldsymbol{\mathcal{E}}(\mathbf{R})\cdot \mathbf{R} + e\bigg(\frac{m^{2}_{\mathrm{c}}}{M^{2}} - \frac{2 m^{2}_{\mathrm{e}}}{M^{2}}\bigg) \boldsymbol{\mathcal{E}}(\mathbf{r})\cdot \mathbf{r} \\
    &+ e\bigg(\frac{m_{\mathrm{c}}}{M} + \frac{2 m_{\mathrm{e}}}{M}\bigg) [\boldsymbol{\mathcal{E}}(\mathbf{r})\cdot \mathbf{R} + \boldsymbol{\mathcal{E}}(\mathbf{r})\cdot \mathbf{R}].
    \end{aligned}
\end{equation}
We proceed by approximating $m_{\mathrm{c}} \approx M$ and discarding corrections of order $m_{\mathrm{e}}/M$, which are typically on the order of $10^{-5}$.
As a result, we obtain the single-ion Hamiltonian in terms of internal, external, and coupling Hamiltonian terms:
\begin{equation}
    \begin{aligned}
    H &= \underbrace{\frac{1}{2m_{\mathrm{e}}} \bigg[\mathbf{p}+ \frac{e}{2} \big(\mathbf{B} \times  \mathbf{r}\big)\bigg]^2 - \boldsymbol{\mu}_{\mathrm{e}}\cdot \mathbf{B} + V(r) + e \boldsymbol{\mathcal{E}}(\mathbf{r})\cdot \mathbf{r}}_{\text{internal dynamics}} + \underbrace{\frac{1}{2M} \bigg[\mathbf{P}- \frac{e}{2} \big(\mathbf{B} \times  \mathbf{R}\big)\bigg]^2 - e \boldsymbol{\mathcal{E}}(\mathbf{R})\cdot \mathbf{R}}_{\text{external dynamics}} \\
    &+ \underbrace{\frac{e}{M} (\mathbf{B}\times \mathbf{r}) \cdot \mathbf{P}- \frac{e^2}{2M}(\mathbf{B}\times \mathbf{R})\cdot (\mathbf{B}\times \mathbf{r}) + e \boldsymbol{\mathcal{E}}(\mathbf{R})\cdot \mathbf{r}+ e \boldsymbol{\mathcal{E}}(\mathbf{r})\cdot \mathbf{R}}_{\text{coupling between internal and external}} = H_{\mathrm{in}}(\mathbf{r}) +H_{\mathrm{ex}}(\mathbf{R}) + H_{\mathrm{co}}(\mathbf{R}, \mathbf{r}),
    \end{aligned}
\end{equation}
which corresponds to Eq.~\eqref{eq:ham_total}.
Here, $H_{\mathrm{ex}}$ describes a free particle with mass $M$ and charge $q = e$, moving in a homogeneous magnetic and quadrupole electric fields; $H_{\mathrm{in}}$ represents a particle of mass $m_{\mathrm{e}}$ and charge $q = - e$, also subjected to homogeneous magnetic and quadrupole electric fields, but additionally influenced by the central potential $V(r)$.
\end{widetext}

\section{Internal Hamiltonian diagonalization and Rydberg spectrum}
\label{app:numerics}

In this appendix, we present the computational approach used to obtain the Rydberg energy spectrum.
In the Paschen--Back regime, the internal Hamiltonian is approximately diagonal in the basis
\begin{equation}
    H_{\mathrm{in}} \approx \sum_{\mathbf{L}} E_{\mathbf{L}}\, \ket{\mathbf{L}}\!\bra{\mathbf{L}},
\end{equation}
where the labels $\mathbf{L}$ correspond to the asymptotic eigenstates at large magnetic field.
To incorporate the $l$-mixing induced by the diamagnetic interaction, we introduce the dressed eigenstates at a finite magnetic field:
\begin{equation}
    H_{\mathrm{in}}(B)\ket{E_{\mathbf{L}}(B)} = E_{\mathbf{L}}(B)\ket{E_{\mathbf{L}}(B)},
\end{equation}
which evolve adiabatically from the Paschen--Back states $\ket{\mathbf{L}}$.
The expectation value of an operator $O$ in this basis is evaluated as
\begin{equation} 
    \bra{E_{\mathbf{L}}} O \ket{E_{\mathbf{L}'}} = \sum_{\mathbf{J}\mathbf{J}'} \braket{\mathbf{J} \!\mid \!E_{\mathbf{L}}}^{\star} \braket{\mathbf{J}' \!\mid \!E_{\mathbf{L}'}} \bra{\mathbf{J}} O \ket{\mathbf{J}'}, 
\end{equation}
with $\ket{\mathbf{J}}=\ket{n,l,j,m_j}$ defined in Eq.~\eqref{eq:fine_basis}.
We have omitted explicit dependence on
the magnetic field strength $B$.

In practice, the quantities $E_{\mathbf{L}}(B)$ and $\ket{E_{\mathbf{L}}(B)}$ are obtained by full numerical diagonalization of $H_{\mathrm{in}}(B)$ in a truncated Hilbert space. 
Since we focus on the Rydberg $S$ and $P$ manifolds, we retain the isolated set of neighboring states
\begin{equation} 
    \begin{aligned} 
    &\ket{(n-2)F_{j}}, \ket{(n-2)G_{j}}, \ket{(n-2)H_{j}}; \\ &\ket{(n-1)D_{j}}, \ket{(n-1)F_{j}}, \ket{(n-1)G}, \ket{ (n-1)H_{j}}; \\ &\ket{nS_{j}}, \ket{nP_{j}}. 
    \end{aligned} 
    \label{eq:iso_states} 
\end{equation}
which results in a $126\times 126$ Hamiltonian matrix for $0 \le l \le 5$ and $\lvert l - 1/2\rvert \le j \le l + 1/2$.
The homogeneous magnetic field and electric quadrupole field generate couplings that satisfy the quadrupole selection rules $\Delta l = 0, \pm 2$.
Consequently, second-order energy shifts of the low-$l$ states ($l = 0, 1, 2$) involve transitions to states with $l \le 4$ (G-states), which justifies the chosen truncation of the Hilbert space.

\section{Electronic losses and ionization due to external fields}
\label{app:ionization}

In this appendix, we derive the thresholds for unwanted electron loss due to external electric and magnetic fields.
To estimate these thresholds, we neglect the spin-orbit coupling and approximate the central potential as
\begin{equation}
    V(r) \approx - \frac{e^{2}}{2\pi \epsilon_{0} r}.
\end{equation}
The resulting internal Hamiltonian can be decoupled into Landau-like levels by expressing the wave function as 
\begin{equation}
    \psi(\rho, z, \phi) = f(\rho, z)\e^{\mathrm{i}m_{l}\phi},
\end{equation}
where $\rho$ is the radial coordinate and $m_{l} = 0, \pm 1, \pm 2, \ldots$ is the magnetic quantum number.
The function $f(\rho, z)$ satisfies the Schrödinger equation $H'_{\mathrm{in}} f(\rho, z) = E f(\rho, z)$, with
\begin{equation}
    \begin{aligned}
        H'_{\mathrm{in}} &= -\frac{1}{2m_{\mathrm{e}}} \bigg[\frac{1}{\rho}\frac{\partial\phantom{x}}{\partial \rho} \bigg(\rho \frac{\partial\phantom{x}}{\partial \rho}\bigg) + \frac{\partial^{2}\phantom{x}}{\partial z^{2}}\bigg] + \frac{eB}{2m_{\mathrm{e}}}m_{l} \\
        &- \frac{e^{2}}{2\pi \epsilon_{0} \sqrt{\rho^{2} + z^{2}}} + \frac{m^{2}_{l}}{2m_{\mathrm{e}} \rho^{2}} + \bigg(\!e\beta + \frac{e^{2}B^{2}}{8m_{\mathrm{e}}}\!\bigg) \rho^{2} - 2e\beta z^{2}.
    \end{aligned}
\end{equation}
We identify the associated effective two-dimensional potential describing the electron confinement in the Hamiltonian $H'_{\mathrm{in}}$:
\begin{equation}
    \begin{aligned}
        V_{\mathrm{eff}}(\rho, z) &= \frac{eB}{2m_{\mathrm{e}}}m_{l} - \frac{e^{2}}{2\pi \epsilon_{0} \sqrt{\rho^{2} + z^{2}}} \\
        &+ \frac{m^{2}_{l}}{2m_{\mathrm{e}} \rho^{2}} + \bigg(e\beta + \frac{e^{2}B^{2}}{8m_{\mathrm{e}}}\bigg) \rho^{2} - 2e\beta z^{2}.
    \end{aligned}
\end{equation}

We now perform a stability analysis of this potential to determine the ionization limit.
Minimizing $V_{\mathrm{eff}}$ with respect to the $z$-coordinate yields the equation
\begin{equation}
    \left.\frac{\partial V_{\mathrm{eff}}}{\partial z}\right|_{z = z_{0}} = \bigg[\frac{e^{2}}{2\pi \epsilon_{0}(\rho^{2} + z^{2})^{3/2}} - 4e\beta\bigg] z_{0} = 0.
\end{equation}
This equation leads to two stationary branches: an axisymmetric extremum at $z_{0} = 0$, and an off-axis saddle point defined by $r_{1}^{3} = (\rho_{1}^{2} + z_{1}^{2})^{3/2} = e/(8\pi \epsilon_{0} \beta)$.

For the first branch ($z_{0} = 0$), the confinement condition $\tfrac{\partial^{2} V_{\mathrm{eff}}}{\partial z^{2}} > 0$ implies the condition $\rho < [e/(8\pi \epsilon_{0} \beta)]^{1/3}$ for the radial coordinate.
For $m_{l} = 0$, which corresponds to the Rydberg states considered in the main text, the absence of the centrifugal term ensures that the potential minimum lies at $\rho = 0$.
For the second branch, the stationary condition with respect to $\rho$ at $r = r_{1}$ gives
\begin{equation}
    \left. \frac{\partial V_{\mathrm{eff}}}{\partial \rho} \right|_{r = r_{1}} = - \frac{m^{2}_{l}}{m{\mathrm{e}} \rho^{3}} + \bigg(6 e\beta + \frac{e^{2}B^{2}}{4m_{\mathrm{e}}}\bigg) \rho = 0,
\end{equation}
which corresponds to a saddle point located at
\begin{equation}
    \rho_{1} = \bigg(\frac{2m^{2}_{l}}{24 e m_{\mathrm{e}}\beta + e^{2}B^{2}}\bigg)^{1/4}.
\end{equation}
For $m_{l} = 0$, we obtain $\rho_{1} = 0$, and the axial coordinate yields two saddle points along the axial direction, $z_{1} = \pm [e/(8\pi \epsilon_{0} \beta)]^{1/3}$.
In addition, the effective potential evaluated at these points is
\begin{equation}
    V_{\mathrm{eff}}(0, z_{1}) = - \frac{3}{2} \bigg(\frac{e^{5}\beta}{\pi^{2}\epsilon_{0}^{2}}\bigg)^{3/2}.
\end{equation}
The ionization electric field gradient is obtained by equating $V_{\mathrm{eff}}(0, z_{1})$ to the bound-state energies approximated to $E_{n} = -2/(m_{\mathrm{e}}a_{0}^{2}n^{2})$.
As a result, $\beta_{\mathrm{ion}} = 9.2 \times 10^{10}\,\text{V/m}^{2}$ for Rydberg states with principal quantum number $n = 50$.
These gradients are almost $10^{4}$ times stronger than the maximum values realized in Penning traps, as discussed in the main text.
Under typical operating conditions, with $\beta = 10^{7}\, \mathrm{V/m}^{2}$, we find the ultimate ionization limit corresponds to the Rydberg state with $n = 228$.

\section{Coupling between internal and external dynamics}
\label{app:coup_intext}

In this appendix, we show the calculations of the second-order energy shifts due to internal-external coupling shown in Eq.~\eqref{eq:shift_energy}.
We start by rewriting the coupling Hamiltonian as
\begin{equation}
    \begin{aligned}
        H_{\mathrm{co}} 
        = &- \sqrt{\frac{2\pi}{3}} r \bigg\{Y^{1}_{1}(\phi, \theta)\bigg[\frac{eB}{M} (P_{y} - \mathrm{i}P_{x}) \\
        &\qquad \qquad \quad+ \Big(2e\beta - \frac{e^{2}B^{2}}{2M}\Big)(X + \mathrm{i}Y)\bigg] + \mathrm{H.c.}\bigg\}\\
        &- \sqrt{\frac{4\pi}{3}}r Y^{0}_{1}(\phi, \theta) 4 e \beta Z
    \end{aligned}
\end{equation}
where $Y^{m_{l}}_{l} (\phi, \theta)$ are spherical harmonics, and $\phi$ and $\theta$ are azimuth and polar angles with respect to the $z$-axis.
In terms of the spherical harmonics, the couplings provoked by the internal-external coupling are made explicit, since terms $\propto Y^{m_{\kappa}}_{1}(\phi, \theta)$ only couple states with $\Delta l = 1$ and $\Delta m_{l} = m_{\kappa}$.

Assuming internal electronic states in the Paschen-Back regime, the second-order energy correction due to the internal-external coupling results in
\begin{equation}
    \Delta E^{(2)}_{\mathbf{L}}(\mathbf{R}, \mathbf{P}) = \sum_{\mathbf{L'} \ne \mathbf{L}} \frac{\left|\bra{\mathbf{L}} H_{\mathrm{co}} \ket{\mathbf{L}'}\right|^{2}}{E_{\mathbf{L}'} - E_{\mathbf{L}}},
\end{equation}
where $\ket{\mathbf{L}'} = \ket{n', l', m'_{l}, m'_{s}}$.
The numerator of this expression is explicitly calculated as
\begin{widetext}
    \begin{equation}
        \begin{aligned}
            &\left|\bra{\mathbf{L}} H_{\mathrm{co}} \ket{\mathbf{L}'}\right|^{2} \\
            &= \frac{\left|\bra{\mathbf{L}}r\ket{\mathbf{L}'}\right|^{2}}{3} \bigg[\frac{e^{2}B^{2}}{M^{2}} \big(P^{2}_{x} + P^{2}_{y}\big) + \Big(2e\beta - \frac{e^{2}B^{2}}{2M}\Big)^{2} (X^{2} + Y^{2}) + \frac{2eB}{M}\Big(2e\beta - \frac{e^{2}B^{2}}{2M}\Big) (L_{z} + 2YP_{x}) \bigg] \delta_{l, l'+1}\delta_{m_{l}, m'_{l}\pm 1} \\
            &+ \frac{32\left|\bra{\mathbf{L}}r\ket{\mathbf{L}'}\right|^{2}}{3} e^{2}\beta^{2}Z^{2} \delta_{l, l'+1}\delta_{m_{l}, m'_{l}}.
        \end{aligned}
    \end{equation}
\end{widetext}
Thus, the expressions for the frequency and mass modifications in terms of the perturbative series are given by
\begin{equation}
    \begin{aligned}
        \tilde{M} &= - \frac{M^{2}}{e^{2}B^{2}} \bigg[\sum_{\mathbf{L}'}\frac{\left|\bra{\mathbf{L}} r \ket{\mathbf{L}'}\right|^{2}}{E_{\mathbf{L}'} - E_{\mathbf{L}}}\bigg]^{-1} \delta_{l, l'+1} \delta_{m_{l}, m'_{l} \pm 1},\\
        \tilde{\omega}^{2}_{\rho}&= - \frac{1}{M}\bigg(2e\beta - \frac{e^{2}B^{2}}{2M}\bigg)^{2}\sum_{\mathbf{L}'}\frac{\left|\bra{\mathbf{L}} r \ket{\mathbf{L}'}\right|^{2}}{E_{\mathbf{L}'} - E_{\mathbf{L}}} \delta_{l, l'+1} \delta_{m_{l}, m'_{l} \pm 1},
    \end{aligned}
\end{equation}
\begin{equation}
    \begin{aligned}
        \tilde{\omega}^{2}_{z} &= - \frac{32e^{2}\beta^{2}}{3M}\sum_{\mathbf{L}'}\frac{\left|\bra{\mathbf{L}} r \ket{\mathbf{L}'}\right|^{2}}{E_{\mathbf{L}'} - E_{\mathbf{L}}} \delta_{l, l'+1} \delta_{m_{l}, m'_{l}} ,\\
        \tilde{\omega}_\mathrm{c} &= -\frac{eB}{3M} \bigg(4 e \beta - \frac{e^{2}B^{2}}{M}\bigg) \sum_{\mathbf{L}'}\frac{\left|\bra{\mathbf{L}} r \ket{\mathbf{L}'}\right|^{2}}{E_{\mathbf{L}'} - E_{\mathbf{L}}} \delta_{l, l'+1} \delta_{m_{l}, m'_{l} \pm 1}.
    \end{aligned}
\end{equation}
Considering the state $\ket{\mathbf{L}}$ to be $\ket{S} = \ket{n, 0, 0, \pm1/2}$, and calculating only the first dipole-allowed transition to the next-neighbor state, $n' = n$, we obtain the expressions in Eq.~\ref{eq:corr_shifts}.

\section{Three-ion crystal vibrational normal modes}
\label{app:normal_modes}

\begin{figure}
    \centering
    \includegraphics[scale=0.4]{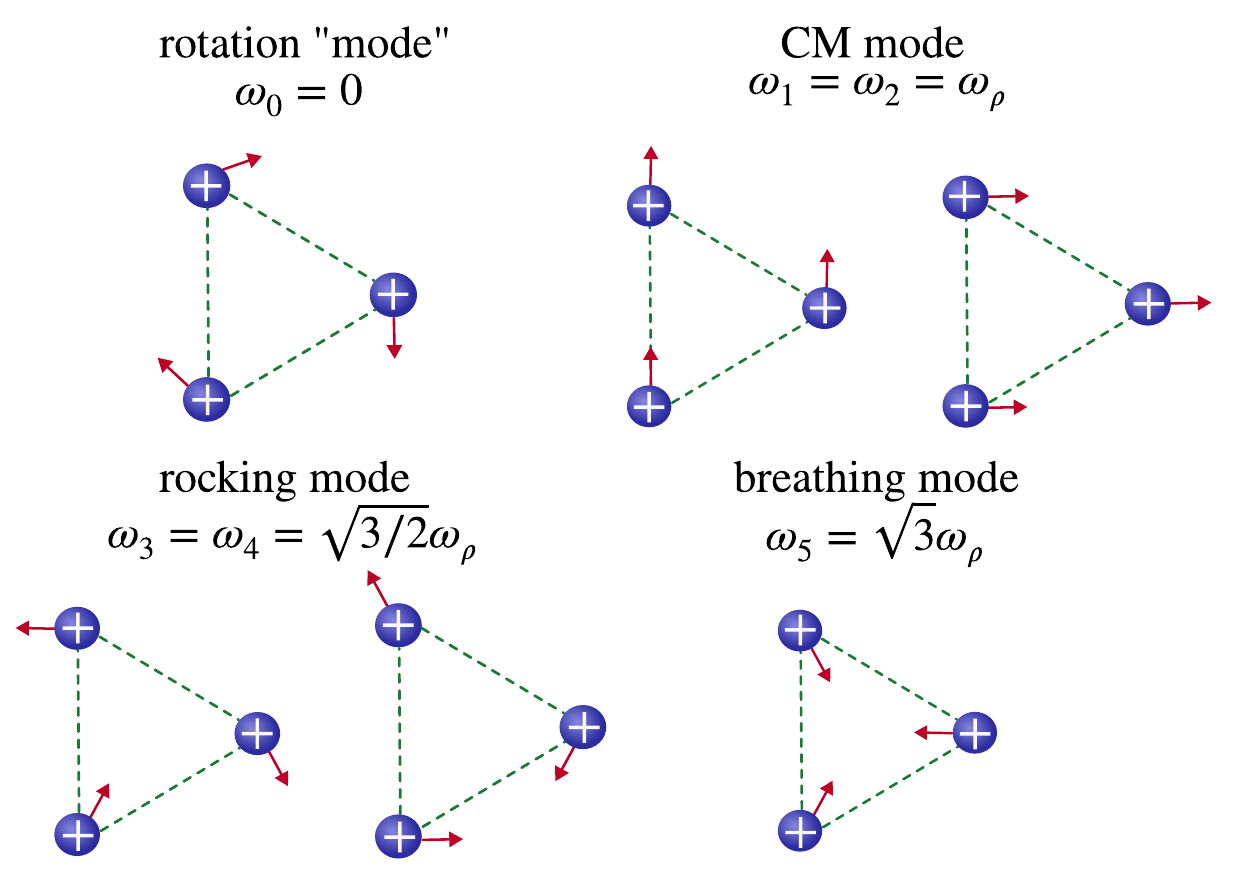}
    \caption{\textbf{Collective vibrational normal modes in the three-ion crystal.}
    The eigenfrequency $\sqrt{3}\omega_{\rho}$ corresponds to the breathing mode, the degenerate eigenfrequencies $\sqrt{3/2}\omega_{\rho}$ and $\omega_{\rho}$ correspond to rocking and CM mode, respectivelly. 
    We identify a vanishing frequency, $\omega_{0}$, associated with the rotation of the ion crystal about the $z$-axis.
    }
    \label{fig:n_modes}
\end{figure}

In this appendix, we derive the vibrational normal modes of the planar three-ion crystal described in the main text. 
The external Hamiltonian reads
\begin{equation}
    \begin{aligned}
        \mathcal{H}_{\mathrm{ex}} &= T_{\mathrm{ex}} + V_{\mathrm{ex}},
    \end{aligned}
    \label{eq:ion_ham}
\end{equation}
where $T_{\mathrm{ex}} = \sum^{3}_{i = 1} \sum_{u = x, y} P^{2}_{u, i}/2M$ is the external kinect energy and
\begin{equation}
    V_{\mathrm{ex}} =  \sum^{3}_{i = 1} \frac{1}{2} M \omega^{2}_{\rho} \big(X^{2}_{i} + Y^{2}_{i}\big)+ \frac{1}{2}\frac{1}{4\pi \epsilon_{0}} \sum^{3}_{\substack{i,j=1\\i \ne j}} \frac{e^{2}}{R_{ij}},
\end{equation}
is the external potential energy.
To obtain the vibrational motion about the potential equilibrium, we proceed by expanding the potential $V_{\mathrm{ex}}$ in terms of small displacements around the equilibrium positions: $\delta \mathbf{R}_{i} = \mathbf{R}_{i} - \mathbf{R}^{0}_{i}$.
We group the small displacement coordinates in a single vector given by $\mathbf{Q} = \mathbf{Q}' - \mathbf{Q}^{0}$, with
\begin{equation}
    \begin{aligned}
         \mathbf{Q}' &= (X_{1}, X_{2}, X_{3}, Y_{1}, Y_{2}, Y_{3})^{\mathrm{T}},\\
        \mathbf{Q}^{0} &= (X^{0}_{1}, X^{0}_{2}, X^{0}_{3}, Y^{0}_{1}, Y^{0}_{2}, Y^{0}_{3})^{\mathrm{T}}.
    \end{aligned}
\end{equation}
Performing the harmonic expansion in the external potential to the second order in the small displacements $\mathbf{Q}$, we obtain
\begin{equation}
    V_{\mathrm{ex}} (\mathbf{Q}) = \frac{M}{2} \mathbf{Q}^{\mathrm{T}} \mathcal{K} \mathbf{Q}.
    \label{eq:quad_pot}
\end{equation}
Here, the matrix $\mathcal{K}$ is the Hessian matrix, defined as $\mathcal{K} = M^{-1}(\nabla \otimes \nabla^{\mathrm{T}})_{\mathbf{q} = \mathbf{q}^{0}} V_{\mathrm{ex}}$ and explicitly given by
\begin{equation}
    \mathcal{K}/\omega^{2}_{\rho} = \left(\!
    \begin{array}{cccccc}
    \frac{11}{6} & -\frac{5}{12}  & -\frac{5}{12}  & 0 & \frac{\sqrt{3}}{4} & -\frac{\sqrt{3}}{4}  \\
    -\frac{5}{12} & \frac{13}{12} & \frac{1}{3} & \frac{\sqrt{3}}{4} & -\frac{\sqrt{3}}{4}  & 0 \\
    -\frac{5}{12} & \frac{1}{3} & \frac{13}{12} & -\frac{\sqrt{3}}{4}  & 0 & \frac{\sqrt{3}}{4} \\
    0 & \frac{\sqrt{3}}{4} & -\frac{\sqrt{3}}{4}  & \frac{5}{6} & \frac{1}{12} & \frac{1}{12} \\
     \frac{\sqrt{3}}{4} & -\frac{\sqrt{3}}{4}  & 0 & \frac{1}{12} & \frac{19}{12} & -\frac{2}{3} \\
    -\frac{\sqrt{3}}{4}  & 0 & \frac{\sqrt{3}}{4} & \frac{1}{12} & -\frac{2}{3} & \frac{19}{12} \\
\end{array}
\!\right).
\end{equation}

\begin{table*}[ht]
\caption{\textbf{One-photon transition properties for different ion species.} 
The table shows ionization mass and ionization specifics \cite{NIST_ASD} for possible ion species for quantum simulation with Rydberg ions in a Penning trap.}
\begin{tabular}{ |c| c| c | c| c| c|} 
   \hline 
   Ion species & \makecell{Mass (most abundant \\ isotopes)} & \makecell{GS ionization \\ limit} & \makecell{One-photon photo- \\ ionization  wavelength} & \makecell{$D_{5/2}$-states ionization \\ limit in wavenumber} & \makecell{$D$-states  one-photon \\ photoionization} \\
  \hline
  $\mathrm{Be}^{+}$ & 9 & $146882 \, \mathrm{cm}^{-1}$ & $68.1\, \mathrm{nm}$  & --- & ---\\ 
  \hline
  $\mathrm{Mg}^{+}$ & 24 & $121268 \, \mathrm{cm}^{-1}$ & $82.5\, \mathrm{nm}$ & --- & ---\\ 
  \hline
  $\mathrm{Ca}^{+}$ & 40 & $95752 \, \mathrm{cm}^{-1}$ & $104.4\, \mathrm{nm}$ & $82041 \, \mathrm{cm}^{-1}$ & $121.9\, \mathrm{nm}$\\ 
  \hline 
  $\mathrm{Sr}^{+}$ & 88 & $88964 \, \mathrm{cm}^{-1}$ & $112.4\, \mathrm{nm}$ & $74128 \, \mathrm{cm}^{-1}$ &  $134.9\, \mathrm{nm}$\\ 
  \hline 
  $\mathrm{Ba}^{+}$ & 138 & $80686 \, \mathrm{cm}^{-1}$ & $123.9\, \mathrm{nm}$ & $75011 \, \mathrm{cm}^{-1}$ &  $133.3\, \mathrm{nm}$\\ 
  \hline 
  $\mathrm{Yb}^{+}$ & 174 & $98207 \, \mathrm{cm}^{-1}$ & $101.8\, \mathrm{nm}$ & --- & ---\\ 
  \hline 
\end{tabular}
\label{tab:ion}
\end{table*}

\begin{table}[ht]
\caption{\textbf{Two-photon transition properties for different ion species.} 
The table shows two-photon excitation schemes and their respective wavelengths for different ion species \cite{NIST_ASD, Heugel_2016}.}
\begin{tabular}{ |c| c| c |} 
   \hline 
   Element & Transitions & Wavelength \\
  \hline
  $\mathrm{Ca}^{+}$ &  \makecell{$3D_{5/2} \rightarrow 5P_{3/2}$ \\ $5P_{3/2} \rightarrow $ cont.} & \makecell{$213.2\, \mathrm{nm}$ \\ $284.6\, \mathrm{nm}$} \\ 
  \hline 
  $\mathrm{Sr}^{+}$ & \makecell{$4D_{5/2} \rightarrow 6P_{3/2}$  \\ $6P_{3/2} \rightarrow$ cont.} &  \makecell{$242.6\, \mathrm{nm}$ \\ $303.9\, \mathrm{nm}$} \\ 
  \hline 
  $\mathrm{Ba}^{+}$ &  \makecell{$5D_{5/2} \rightarrow 7P_{3/2}$ \\ $7P_{3/2} \rightarrow$ cont.} &  \makecell{$225.5\, \mathrm{nm}$ \\ $326.0\, \mathrm{nm}$} \\ 
  \hline 
  $\mathrm{Yb}^{+}$ & \makecell{ $5D_{3/2} \rightarrow 7P_{1/2}$ \\  $7P_{1/2} \rightarrow$ cont.}  &   \makecell{$245.4\, \mathrm{nm}$ \\ $289.8\, \mathrm{nm}$} \\ 
  \hline 
\end{tabular}
\label{tab:2ph}
\end{table}

The vibrational normal modes are obtained by solving the eigenproblem associated with the matrix $\mathcal{K}$, i.e., 
\begin{equation}
    \mathcal{M}^{-1} \mathcal{K}\mathcal{M} = \mathbf{\Omega}^{2},
\end{equation}
where 
\begin{equation}
        \mathbf{\Omega} = \mathrm{diag} (\omega_{0}, \omega_{1}, \omega_{2}, \omega_{3}, \omega_{4}, \omega_{5}),
\end{equation}
where $\mathcal{M}$ is the normal mode matrix and $\mathbf{\Omega}$ is the matrix of normal mode frequencies.
Explicitly, the normal mode matrix is written as 
\begin{equation}
    \mathcal{M} = \left(\!
    \begin{array}{cccccc}
    0 & \frac{1}{\sqrt{3}} & 0 & -\frac{1}{2 \sqrt{3}} & \frac{1}{2} & -\frac{1}{\sqrt{3}} \\
    \frac{1}{2} & \frac{1}{\sqrt{3}} & 0 & -\frac{1}{2 \sqrt{3}} & -\frac{1}{2} & \frac{1}{2 \sqrt{3}} \\
    -\frac{1}{2} & \frac{1}{\sqrt{3}} & 0 & \frac{1}{\sqrt{3}} & 0 & \frac{1}{2 \sqrt{3}} \\
    -\frac{1}{\sqrt{3}} & 0 & \frac{1}{\sqrt{3}} & -\frac{1}{2} & -\frac{1}{2 \sqrt{3}} & 0 \\
    \frac{1}{2 \sqrt{3}} & 0 & \frac{1}{\sqrt{3}} & \frac{1}{2} & -\frac{1}{2 \sqrt{3}} & -\frac{1}{2} \\
    \frac{1}{2 \sqrt{3}} & 0 & \frac{1}{\sqrt{3}} & 0 & \frac{1}{\sqrt{3}} & \frac{1}{2} \\
\end{array}
\!\right).
\end{equation}
Each column of this matrix represents a vibrational normal mode, which is also a normalized eigenvector of $\mathcal{K}$.
The normal mode frequencies obtained by solving this eigenproblem are 
\begin{equation}
    \omega_{\alpha} \in \{0, \omega_{\rho}, \omega_{\rho}, \sqrt{3/2}\omega_{\rho}, \sqrt{3/2}\omega_{\rho}, \sqrt{3}\omega_{\rho}\}.
    \label{eq:nm_freq}
\end{equation}
Each mode and its respective frequency and degeneracy are shown in Fig.~\ref{fig:n_modes}. 
The first frequency is a zero-mode frequency, i.e., $\omega_{0} = 0$, describing a free particle with no restoring force.
This is due to the three-ion crystal symmetry, which allows free rotations about the $z$ direction.

To diagonalize the external Hamiltonian, we introduce a set of annihilation and creation operators $a_{\alpha}$ and $a^{\dagger}_{\alpha}$ of each phonon mode, totaling five modes.
This allows us to write the canonical coordinates and conjugate momenta as
\begin{equation}
    \begin{aligned}
    \mathbf{Q}_{\beta} &= \sum^{5}_{\alpha = 1} \mathcal{M}_{\beta \alpha} \ell_{\alpha} (a^{\dagger}_{\alpha} + a_{\alpha}), \\
    \mathbf{P}_{\beta} &= \sum^{5}_{\alpha = 1} \mathcal{M}_{\beta \alpha} \mathrm{i}\wp_{\alpha} (a^{\dagger}_{\alpha} - a_{\alpha}),
    \end{aligned}
\end{equation}
where we have introduced characteristic lengths and momenta, $\ell_{\alpha} = \sqrt{1/(2M\omega_{\alpha}})$ and $\wp_{\alpha} = \sqrt{( M\omega_{\alpha})/2}$, respectively.
In terms of $a_{\alpha}$ and $a^{\dagger}_{\alpha}$, the external Hamiltonian assumes the diagonal form 
\begin{equation}
    \mathcal{H}_{\mathrm{ex}} = \sum^{5}_{\alpha = 1} \omega_{\alpha} \big(a^{\dagger}_{\alpha}a_{\alpha} + 1/2\big).
\end{equation}

\section{Different ion species}
\label{app:ion_species}

In this appendix, we present the electromagnetic properties and excitation schemes of different ion species that can be used for Rydberg excitation in a Penning trap. 
Table~\ref{tab:ion} summarizes the properties relevant to one-photon excitation for six ion species.
We note that $\mathrm{Be}^{+}$ and $\mathrm{Mg}^{+}$ do not possess a metastable $D$ state and therefore require more elaborate excitation schemes, typically involving vacuum-ultraviolet lasers or multi-step excitation. 
In contrast, the remaining ion species can be excited via two-photon schemes, as demonstrated experimentally for $\mathrm{Ca}^{+}$ and $\mathrm{Sr}^{+}$ ions~\cite{bao_vo_2025, thomm_2026}. 
The corresponding two-photon excitation pathways and their associated wavelengths are listed in Table~\ref{tab:2ph}.

\section{Dephasing due to fluctuating fields}
\label{app:fluc_fields}

In this appendix, we estimate the dephasing of the effective spin degrees of freedom induced by fluctuations of the MW and magnetic fields.
Here, the spin states correspond to the dressed electronic states introduced in Eq.~\eqref{eq:detuning}, and we focus on the decay of their quantum coherence.

The Hamiltonian in Eq.~\eqref{eq:spin_ham} shows that such fluctuations affect both the single-particle term proportional to $\Delta$ and the interaction term proportional to $V_{0}$.
The resulting dephasing, in addition to the contribution from the finite lifetime of the dressed Rydberg state, acquires a modulation that reads
\begin{equation}
    f(t) = \overline{e^{-i\varphi(t)}},
\end{equation}
which describes the ensemble-averaged decay of the off-diagonal element of the spin density matrix~\cite{monz_2011, monz_2011b}.
The accumulated random phase is
\begin{equation}
    \varphi(t) = \int^{t}_{0} \mathrm{d}t'\,\delta F(t')
\end{equation}
where $\delta F$ denotes the fluctuation of a parameter $F$, and
$\overline{\phantom{l}\circ\phantom{l}}$
represents an average over noise realizations~\cite{monz_2011}.
For fluctuations of the magnetic field $B$ and MW Rabi frequency $\Omega_{\mathrm{MW}}$, the induced fluctuation of $F$ can be expressed to first order as
\begin{equation}
    \delta F = \frac{\partial F}{\partial B}\,\delta B 
    +
    \frac{\partial F}{\partial \Omega_{\mathrm{MW}}}\,\delta\Omega_{\mathrm{MW}}.
    \label{eq:fluc_F}
\end{equation}
Here, we assume that
$\delta B$ and
$\delta\Omega_{\mathrm{MW}}$ are Gaussian distributed and constant during a single experimental realization.
Using $ \overline{e^{-\mathrm{i}\varphi(t)}} = e^{-\overline{\varphi^2(t)}/2}$, for zero mean Gaussian fluctuations, one obtains
\begin{equation} 
    \begin{aligned} 
    f(t) &= \exp \bigg\{ - \frac{1}{2}\bigg[\bigg(\frac{\partial F}{\partial B}\bigg)^{2}\sigma^{2}_{B} + \bigg(\frac{\partial F}{\partial \Omega_{\mathrm{MW}}}\bigg)^{2} \sigma^{2}_{\mathrm{MW}}\bigg]t^{2}\bigg\} \\ &= \e^{-\big(\Gamma^{2}_{B} + \Gamma^{2}_{\mathrm{MW}}\big)t^{2}}. 
    \end{aligned} 
\end{equation}
Here, $\sigma_B = \big(\overline{\delta B^2}\big)^{\frac{1}{2}}$ and
$\sigma_{\mathrm{MW}} = \big(\overline{\delta\Omega_{\mathrm{MW}}^2}\big)^{\frac{1}{2}}$ denote the root-mean-square amplitudes of the magnetic- and MW-field fluctuations, respectively.
The quantities $\Gamma_B$ and $\Gamma_{\mathrm{MW}}$ characterize the corresponding Gaussian dephasing.

The dephasing induced by fluctuations of the detuning $\Delta$ can be obtained from Eq.~\eqref{eq:fluc_F}.
The derivatives are obtained from Eq.~\eqref{eq:detuning}, resulting in
\begin{equation}
        \frac{\partial \Delta}{\partial B}
        =
        \frac{1}{2}
        \bigg(
        \frac{\partial E_{S}}{\partial B}
        +
        \frac{\partial E_{P}}{\partial B}
        \bigg),
        \qquad
        \frac{\partial \Delta}{\partial \Omega_{\mathrm{MW}}}
        =
        -\frac{1}{2},
        \label{eq:delta_db}
\end{equation}
which are valid for resonant dressing, i.e., $\Delta_{\mathrm{MW}}=0$.
The magnetic-field contribution is determined by the field dependence of the dressed-state energies $E_S(B)$ and $E_P(B)$, whereas the sensitivity to MW-amplitude fluctuations is constant.

In addition, magnetic-field fluctuations induce fluctuations of the interaction strength $V_0$.
Using Eq.~\eqref{eq:eq_position}, the magnetic-field dependence of $V_0$ can be obtained explicitly.
For resonant dressing, $V_0$ is independent of $\Omega_{\mathrm{MW}}$ and therefore only magnetic-field fluctuations contribute.
The resulting derivative is
\begin{equation}
     \frac{\partial V_{0}}{\partial B}
     =
     \frac{2M}{3}
     \bigg(
     d_z^2 \omega_{\rho}
     \frac{\partial \omega_{\rho}}{\partial B}
     +
     \omega_{\rho}^{2}
     d_z
     \frac{\partial d_z}{\partial B}
     \bigg),
     \label{eq:v0_db}
\end{equation}
where
\begin{equation}
    d_z(B)=
    \left|
    \bra{E_S} z \ket{E_P}
    \right|(B)
\end{equation}
denotes the magnetic-field-dependent dipole matrix element.
Eq.~\eqref{eq:v0_db} reveals two distinct contributions to interaction-strength fluctuations: variations of the radial trapping frequency $\omega_{\rho}$ and variations of the dipole matrix element $d_z$.
The latter becomes dominant at larger magnetic fields, where the diamagnetic interaction increasingly mixes states with different orbital angular momenta [see Fig.~\ref{fig:mw_scheme}(c)].
In contrast, the radial-frequency contribution grows approximately linearly with the magnetic field because $\omega_{\rho}\propto B$.

\begin{figure}[ht]
    \centering
    \includegraphics[scale=0.39]{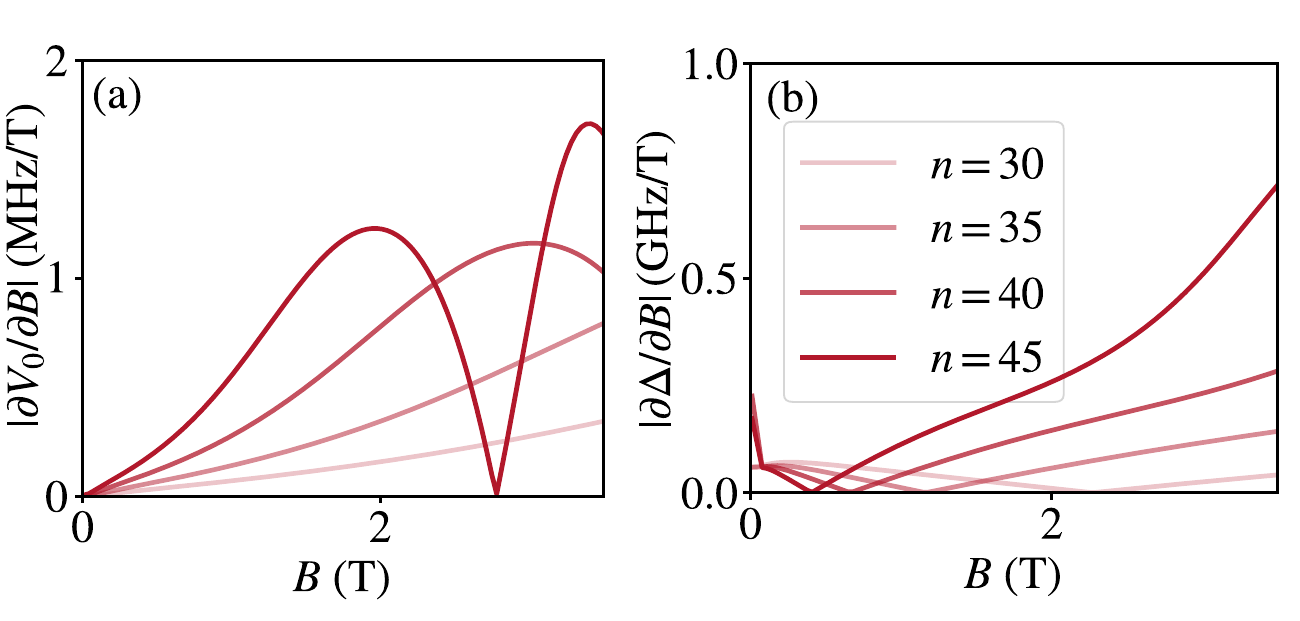}
    \caption{\textbf{Response of dipole-dipole interaction strength and detuning to the magnetic field.}
    Figs.~(a) and (b) show the absolute values of the gradients of the dipole-dipole interaction strength and the detuning, respectively, as a function of the magnetic field for different principal quantum numbers $n$.
    }
    \label{fig:VdB}
\end{figure}

We consider the parameter regime of the experiment presented in Ref.~\cite{zhang_po_wei_2020}.
Active stabilization of the MW power intensity may reduce fluctuations to below $0.01\,\%$.
Since $\Omega_{\mathrm{MW}}\propto \sqrt{P_{\mathrm{MW}}}$, where $P_{\mathrm{MW}}$ denotes the MW power, this corresponds to a relative MW-amplitude noise
\begin{equation}
    \frac{\sigma_{\mathrm{MW}}}{\Omega_{\mathrm{MW}}}
    =
    5\times10^{-5}.
\end{equation}
Furthermore, Penning-trap experiments have demonstrated magnetic-field instabilities below
\begin{equation}
    \frac{\sigma_B}{B}
    <
    2\times10^{-8},
\end{equation}
for averaging times of approximately one minute, even in the absence of magnetic shielding and using only passive thermal stabilization~\cite{mcmahon_2022}.

Using these experimentally reported noise levels together with the derivatives in Eqs.~\eqref{eq:delta_db} and \eqref{eq:v0_db}, we estimate the dephasing strengths quoted in the main text (see Tab.~\ref{tab:decay}).
The absolute values of the relevant derivatives are shown in Fig.~\ref{fig:VdB}(a,b).
These results indicate that dephasing arising from fluctuations of the interaction strength is negligible compared to that caused by fluctuations of the two-level splitting $\Delta$.
For a magnetic field of $B = 2\,\mathrm{T}$, we obtain a dephasing strength $\Gamma_{B} = 2\pi \times 1.1\,\mathrm{kHz}$.
Using the MW Rabi frequency reported in Ref.~\cite{zhang_po_wei_2020}, $\Omega_{\mathrm{MW}} = 2\pi \times 178\,\mathrm{MHz}$,
the corresponding MW-noise amplitude is
$\sigma_{\mathrm{MW}} = 2\pi\times 8.9\,\mathrm{kHz}.$
This yields a dephasing $\Gamma_{\mathrm{MW}} = 2\pi\times 3.0\,\mathrm{kHz}$.

\end{document}